\documentclass{article}       
\usepackage{graphicx}
\usepackage{color,verbatim}
\usepackage{lscape}
\usepackage{rotating}
\usepackage{wrapfig}
\usepackage{multirow}
\usepackage{placeins}
\usepackage{amsmath,bm,natbib}
\usepackage{amsfonts}
\begin{document}

\title{Privacy-preserving estimation of an optimal individualized treatment rule : A case study in maximizing time to severe depression-related outcomes}

\author{Erica EM Moodie, Janie Coulombe, Coraline Danieli, Christel Renoux, \\ and Susan M Shortreed
}

\maketitle

\begin{abstract}
Estimating individualized treatment rules –- particularly in the context of right-censored outcomes -– is challenging because the treatment effect heterogeneity of interest is often small, thus difficult to detect. While this motivates the use of very large datasets such as those from multiple health systems or centres, data privacy may be of concern with participating data centres reluctant to share individual-level data. In this case study on the treatment of depression, we demonstrate an application of distributed regression for privacy protection used in combination with dynamic weighted survival modelling (DWSurv) to estimate an optimal individualized treatment rule whilst obscuring individual-level data. In simulations, we demonstrate the flexibility of this approach to address local treatment practices that may affect confounding, and show that DWSurv retains its double robustness even when performed through a (weighted) distributed regression approach. The work is motivated by, and illustrated with, an analysis of treatment for unipolar depression using the United Kingdom's Clinical Practice Research Datalink. \\ \ \\

\end{abstract}
\ \\ \ \\

\section{Introduction}
\label{sec:intro}

It is widely accepted that different patients respond differently to identical treatments in many clinical settings, particularly the care of depression, motivating the need for individualized treatment rules (ITRs) \citep{Simon2001,Chakraborty2011,aje}. Consider, for example, unipolar depression: selective serotonin reuptake inhibitors (SSRIs), a class of antidepressant drugs, are currently recommended as the first-line treatment option for unipolar depression \citep{ADguidelines}, with citalopram and fluoxetine widely prescribed in several countries \citep[e.g.][]{SSRIsUS,SSRIsUK}. Both drugs have been shown to have similar average effectiveness \citep{CompEff2009}, and so understanding whether there may be a benefit to prescribing one over the other for some patients could be a valuable addition to current clinical guidance. In this work, we are motivated by precisely this question: can we estimate a treatment rule that chooses between two commonly prescribed frontline antidepressants, citalopram and fluoxetine, to maximize the time to severe depression-related outcomes such as hospitalization or suicide in patients with depression? This is accomplished using data from the United Kingdom's Clinical Practice Research Datalink (CPRD) \citep{cprd}, a large primary care database of electronic medical records.

An ITR is a treatment protocol that accounts for treatment effect heterogeneity between people, using patient-level health data to help clinicians tailor treatments to meet individual patient needs \citep{Murphy2003,Robins2004,ChakrabortyMoodie_book}. ITRs are a subset of the larger field of dynamic treatment regimes or adaptive treatment strategies, an area which formalizes the concept of precision medicine in the context of sequential treatment decisions. An ITR focuses on a single treatment decision rather than a sequence of decisions taken based on evolving patient characteristics. One of the main classes of estimation approaches of dynamic treatment regimes is regression-based, an approach which \textit{indirectly} estimates an optimal treatment strategy by first modelling the expected outcome as a function of treatment and covariates and their interactions. Regression-based methods include Q-learning \citep{Watkins:1989, Sutton_Barto:1998}, G-estimation \citep{Robins2004}, and dynamic weighted ordinary least squares (dWOLS) \citep{wallace2015doubly}, as well as generalizations of these to various outcome or exposure types \citep[e.g.][]{goldberg2012q,Rich2016,SimoneauJASA2020,Schulz2021}.

While ITRs are growing part of modern healthcare provision, developing optimal ITRs, i.e.~an ITR that maximizes a patient's expected health outcome, remains challenging, as inference for treatment effect heterogeneity can suffer from low power \citep{Greenland1983}. This is particularly true of censored outcomes, where the number of observed events rather than the number of observed individuals drives standard errors. This can motivate multi-centre collaborations to increase the number of records available for analysis. Such multi-centre collaborations, while promising, bring new challenges, including concerns over possible risks associated with sharing sensitive information \citep{Kaufman2009,Mazor2017} and analytic challenges that may arise if not all centres have collected data on the same covariates. 
In the case of the CPRD, data are available from ten regions of the United Kingdom, where patient case-mix (covariate distributions) and treatment prescribing patterns may vary across these regions or `centres'.

Several methods have been proposed for analyses in distributed data settings, in which individual-level data are not shared with a site conducting the analyses, but remain located (i.e.~distributed) at each site, including data aggregation and meta-analysis \citep{SentinelDataShare}. Data aggregation is suitable for summary statistics, but not for regression analyses that require individual-level data \citep{SentinelDataShare}. A classic meta-analytic approach would appear to be a straightforward way of combining information from multiple settings. However, it has been suggested that meta-analyses are not appropriate in the presence of heterogeneity of treatment effects \citep{GrecoEtAl2013}, which is precisely the setting targeted by ITR analyses, which aim to exploit effect modification of interventions. In addition to concerns about whether a meta-analytic approach to combining ITRs from different data centres would have reasonable statistical properties, the approach may prove impractical, as it requires each regional data centre to have the sufficient funding and statistical expertise needed to perform an ITR analysis. More recently, virtual data pooling \citep{KarrEtAl2007,PSC2017} has been proposed as a promising avenue for combining distributed data without sharing of individual-level information, however it has been demonstrated that double robustness of ITR analyses cannot be assured when applied to pooled data \citep{Danieli2021}.

There is a growing interest in distributed regression \citep{CookSentinelReport2012,CookGrpSeq2014,CookGrpSeq2019,TohClinPharma2020,TohPedRes2020} and other high-level data aggregation or summaries of risks tables in the context of censored outcomes \citep{ShuPrivacy2020} to ensure privacy is maximized while information loss is minimized. A distributed regression approach has been previously shown to be unbiased and to provide consistent, doubly-robust estimators in the context of a regression-based approach to ITR estimation \citep{Danieli2021}. In this work, we adapt this approach to a censored outcome setting and apply it to the CPRD data to estimate an ITR that seeks to maximize the time to severe depression-related outcomes (SDO).

This article is organized as follows. In Section \ref{sec:methods}, the methods including notation and assumptions are detailed. 
We then, in Section \ref{sec:sims}, explore alternative approaches to the distributed regression via simulation, investigating whether there are gains to be had by making ``local'', centre-specific choices (e.g., in covariates needed for confounder adjustment) or whether a ``global'' approach to estimation that requires individual-level data sharing is superior. We apply local and global approaches to the CPRD in Section \ref{sec:cprd}, and conclude with limitations and directions for future investigations in Section \ref{sec:discuss}.

\section{Methods}
\label{sec:methods}

\subsection{Preliminaries: Notation and Assumptions}
Let $i$ index a sample of $n$ individuals, and let $\boldsymbol{X}_i$ be a vector of pre-treatment covariates. We focus on a binary treatment setting, with $A_i \in \{0,1\}$, although extensions to multi-valued or continuous treatments are possible \citep{Schulz2021}. Let $Y_i$ be the time from origin to event, and $T_i$ the observation time for those individuals who are censored, and thus, who are not observed to have the event of interest. We define the counterfactual survival time $Y_i^{a}$ as the potential survival time if, possibly contrary to fact, an individual received treatment $a \in \{0,1\}$. Let $\Delta_i$ denote the event indicator, with realized value denoted $\delta$, which takes the value 1 if the event is observed (and hence $Y_i$ is also observed), and 0 if the subject is censored before the event is observed.
Unless specified otherwise, upper cases, lower cases, and bold respectively denote random variables, realizations of random variables, and vectors.

An ITR is a decision rule $d(\boldsymbol{X}): \boldsymbol{X} \rightarrow \{0,1\}$ that takes as arguments pre-treatment covariates and returns a recommended treatment. An optimal ITR is the decision rule $d^{\text{opt}}(\boldsymbol{x})$ that maximizes the expected log-survival time $\mathbb{E}(\log Y^{a})$ or, assuming that the log-transformation has made the survival time distribution symmetric, maximizes the median survival time such that Med$(\log Y^{d^{\text{opt}}(\boldsymbol{x})}) \ge $ Med$(\log Y^{d})$ for all $d(X) \in \mathcal{D}(X)$ where $\mathcal{D}(X)$ denotes the class of all possible treatment rules. Of course, if the event of interest were something that improves quality of life, such as remission, the methods could easily be adapted such that the ITR minimizes time to the event.

To identify an optimal ITR,  we suppose the axiom of consistency linking counterfactual to actual outcomes, and make the following additional assumptions: (i) no interference, such that an individual's outcome is not influenced by others' treatment allocation, (ii) sequential ignorability \citep{robins2000robust}, i.e.~no unmeasured confounding, and (iii) coarsening at random \citep{gill1997coarsening}, i.e.~the probability of censoring is conditionally independent of the event time, given covariates.

Linking this notation to our motivating example, $Y$ is the time from treatment initiation to a severe SDO, and treatment $A$ is one of two active treatments, citalopram or fluoxetine. The covariates $\boldsymbol{X}$ include demographic variables such as age, sex, as well as health measures such as the number of psychiatric hospital admissions.

\subsection{Dynamic weighted survival modelling}
\label{sec:2}
Dynamic weighted survival modelling (Simoneau et al., \citeyear{SimoneauJASA2020,SimoneauAJE2020,SimoneauRpkg2020})
is a generalization of dynamic weighted ordinary least squares \citep{wallace2015doubly}, a doubly robust regression-based method of estimating adaptive treatment strategies. To proceed, we first define the treatment contrast through the \textit{blip} function:
\begin{equation}
\gamma(a, \boldsymbol{x})=\mathbb{E}[\log(Y^{a})-\log(Y^{0})|\delta=1,\boldsymbol{X}=\boldsymbol{x}], \nonumber
\end{equation}
which is the difference between the expected counterfactual log-survival times of an individual in a world where no one is censored. The blip function must satisfy $\gamma(0, \boldsymbol{x}) = 0$. The optimal individualized treatment rule can then be defined as that which maximizes the blip $a^{\text{opt}} = \text{arg max}_{a} \gamma(a, \boldsymbol{x})$, where $a^{\text{opt}}$ is a function of covariates and thus is a tailored approach to treatment decision-making.

Dynamic weighted survival modelling (DWSurv) relies on an accelerated failure time (AFT) model, as this approach is focused on a scale of direct clinical interest: the (log) survival time. We assume an AFT model for $\log(Y)$:
\begin{equation}
\label{aft1general}
\log(Y^{a})= f(\boldsymbol{x};\boldsymbol{\beta})+ a\cdot g(\boldsymbol{x};\boldsymbol{\psi}) + \epsilon,
\end{equation}
where the errors, $\epsilon$, are independent and identically distributed with $\mathbb{E}(\epsilon)=0$. The model in Equation \ref{aft1general} is separated into two components: a \textit{treatment-free} model $f(\boldsymbol{x};\boldsymbol{\beta})$ for any function $f$ which depends on covariates $\boldsymbol{x}$ but does not depend on treatment, and a treatment effect model $a\cdot g(\boldsymbol{x};\boldsymbol{\psi})$ for any function $g$ which depends on the covariates that are the tailoring variables, which may be a subset of those variables included in $f$. The treatment effect model also includes the main effect of treatment, $a$. Note that there is a slight abuse of notation, in that the subset of $\boldsymbol{x}$ included in $g$ must be contained within the subset included in $f$ but they need not be identical. In particular, we take $a \cdot g(\boldsymbol{x};\boldsymbol{\psi})$ to be a model for $\gamma(a, \boldsymbol{x})$. A typical choice of parameterization is a linear function
\begin{equation}
\label{aft1}
\log(Y^{a})= \boldsymbol{\beta^T}\boldsymbol{x}+ \boldsymbol{\psi^T}a\boldsymbol{x} + \epsilon.
\end{equation}
This then implies
\begin{align}
\gamma(a, \boldsymbol{x};\boldsymbol{\psi})&=\mathbb{E}[\log(Y^{a})-\log(Y^{0})|\delta=1,\boldsymbol{X}=\boldsymbol{x}; \boldsymbol{\psi}] \nonumber\\
&=\boldsymbol{\psi^T}a\boldsymbol{x}. \nonumber
\end{align}
The optimal individualized treatment rule is then $a^{\text{opt}} = I(\boldsymbol{\psi^T}\boldsymbol{x} > 0)$, such that under the above model, we consider the (restricted) class of linear decision rules indexed by the vector of parameters $\boldsymbol{\psi}$. The linear parameterization is not a requirement of the DWSurv approach, however its adoption provides ease of estimation and interpretable decision rules.

\citet{SimoneauJASA2020} showed that estimation of the parameters $\boldsymbol{\psi}$ can be accomplished via a weighted regression. The necessary weights are a product of inverse probability of censoring weights \citep{robins1992recovery} and weights that create balance between the treatment groups. 
In particular, the weights, denoted $w = w(\Delta,A,\boldsymbol{X})$, must satisfy the \textit{balancing property} \citep{SimoneauJASA2020}:
\begin{eqnarray}
\label{balancing}
[1-\varphi(0,\boldsymbol{x})][1-\pi(\boldsymbol{x})]w(0,0,\boldsymbol{x}) &=& \varphi(0, \boldsymbol{x})[1-\pi(\boldsymbol{x})]w(0,1,\boldsymbol{x})\nonumber\\
=[1-\varphi(1,\boldsymbol{x})]\pi(\boldsymbol{x})w(1,0,\boldsymbol{x})&=&\varphi(1,\boldsymbol{x})\pi(\boldsymbol{x})w(1,1,\boldsymbol{x}),
\end{eqnarray}
where $\pi(\boldsymbol{x}) = \mathbb{P}(A=1|\boldsymbol{X}=\boldsymbol{x})$, is the propensity score, and $\varphi(a,\boldsymbol{x}) = \mathbb{P}(\Delta=1|\boldsymbol{X}=\boldsymbol{x},A=a)$, the probability of observing the event of interest, i.e. not being censored. The estimation procedure requires a positivity assumption such that $P(A = a|\boldsymbol{X}) > 0$ for $a \in \{0,1\}$ and all $\boldsymbol{x}$ and $\mathbb{P}(\Delta = 1|\boldsymbol{X},A) > 0$ for all $\boldsymbol{x}$. A familiar choice for these balancing weights is the product of the inverse probability of treatment weights and the inverse probability of censoring weights; \citet{SimoneauJASA2020} also showed that the product of `absolute value' \citep{wallace2015doubly} weights (also called overlap weights \citep{li2018balancing}') for treatment multiplied by the inverse probability of censoring weights also satisfy the balancing equations.

The steps below ensure consistent estimation of the parameters $\boldsymbol{\psi}$ of the individualized treatment rule under the assumption that either the treatment (i.e.~propensity score) and censoring models are correctly specified, or the treatment-free component of the AFT model is correctly specified:
\begin{enumerate}
\item Specify models for each of the probability of treatment and the probability of censoring, $\mathbb{P}(A=1|\boldsymbol{X};\boldsymbol{\alpha})$ and $\mathbb{P}(\Delta=0|\boldsymbol{X}, A;\boldsymbol{\lambda})$.
\item Specify weights $w(\delta, a,\boldsymbol{x};\boldsymbol{\hat\alpha}, \boldsymbol{\hat\lambda})$ that satisfy the balancing property described in Equation \ref{balancing}.
\item Estimate the AFT parameters, $(\boldsymbol{\beta}, \boldsymbol{\psi})$, by solving the following weighted generalized estimating equations \citep{LZgee1986,ZLgee1986}:
\begin{equation}
\label{AFTEstFunc}
U(\boldsymbol{\psi}, \boldsymbol{\beta}) = \sum_{i=1}^n \delta_i \hat w_{i} \begin{pmatrix} \boldsymbol{x_i} \\
a_{i}\boldsymbol{x} \end{pmatrix}\left(\log(Y_i)-\boldsymbol{\beta^T}\boldsymbol{x_i}-\boldsymbol{\psi^T}a_{i}\boldsymbol{x_{i}}\right)=0.
\end{equation}
\end{enumerate}
Note that Equation \ref{AFTEstFunc} takes the form of a weighted regression -- under the linear parameterization of the AFT in Equation \ref{aft1}, a weighted linear regression using outcomes only from those whose events times were observed. We remind the reader that the solution to an estimating function with matrix form $\boldsymbol{X}{^T}W(\tilde{Y} - \boldsymbol{X}\boldsymbol{\theta})$ is given by $\hat{\boldsymbol{\theta}} = (\boldsymbol{X}{^T}W\boldsymbol{X})^{-1}\boldsymbol{X}{^T}W\tilde{Y}$, where, in the context of Equation \ref{AFTEstFunc}, $\boldsymbol{\theta}^{T} = (\boldsymbol{\beta}^T,\boldsymbol{\psi}^T)$, $W$ is the diagonal matrix of balancing weights incorporating both the treatment and censoring probabilities, and $\tilde{Y} = \log(Y)$ is the vector of log-transformed event times. The utility of this matrix formulation will be apparent in the following section.

\subsection{Distributed regression to preserve privacy}
Distributed regression has been proposed as a method both to preserve privacy \citep{TohClinPharma2020,TohPedRes2020} and to compress high-volume data \citep{LuoCJS2021}. It has also been shown to be an effective method of ensuring data protection in the context of ITR analysis of continuous outcomes \citep{Danieli2021}. Suppose there are $J$ centres contributing independent data to the analysis of an ITR, it is then straightforward to show that Equation \ref{AFTEstFunc} can be equivalently expressed as
\[
\hat{\boldsymbol{\theta}} = \left(\sum_{j=1}^J\boldsymbol{X}_j{^T}W_j\boldsymbol{X}_j\right)^{-1}\left(\sum_{j=1}^J\boldsymbol{X}_j{^T}W_j\tilde{Y}_j,
\right) \]
where $\boldsymbol{X}_j$, $W_j$, and $\tilde{Y}_j$ are, respectively, the centre-specific design matrix, diagonal weight matrix, and vector of log-transformed outcomes for each of the $j=1,...,J$ centres. In the context of the CPRD analysis, data are clustered into ten geographic regions (North East, East Midlands, London, and so on) which we treat as ten distinct centres.

Data privacy can be ensured if each centre provides only the two matrices $\boldsymbol{X}_j{^T}W_j\boldsymbol{X}_j$ and $\boldsymbol{X}_j{^T}W_j\tilde{Y}_j$ to a central analysis site, rather than the individual component matrices. This does requires each centre to estimate treatment (i.e.~propensity score) and censoring models independently, which could be accomplished using a set of variables common to all centres or using different models that account for possibly different treatment allocation or censoring mechanisms at each centre.


This approach, which shares summary information from each site, in the form of two matrices is different than the current status quo of multi-site analytic work, in which all data (i.e.~the individual matrix components, $\boldsymbol{X}_j$, $W_j$, and $\tilde{Y}_j$) from each site are shared with the central data analysis site. The central analysis site would then estimate a single treatment model, and a single censoring model, which thereby dramatically reduces (by up to $J$-fold) the number of nuisance parameters to be estimated. This approach may be more statistically efficient, and potentially more pragmatically efficient, if statistical programming resources are scarce. Note, though, that this process assumes that the treatment allocation and censoring processes were identical at each centre; if this is not the case, and site-specific interactions are included in these models, then gains in statistical efficiency are reduced. Furthermore, this ``global'' approach to estimation necessitates sharing of individual patient data, such that data are not held private behind the firewalls of the centre in which they were collected.

In the section that follows, we examine the impact on estimation of the local versus global estimation strategies, and demonstrate that the double robustness of DWSurv is maintained even when combined with the privacy-preserving distributed regression approach to estimation.

\section{Simulations}
\label{sec:sims}

The simulation study is reported following the ADEMP (aims, data-generating mechanisms, estimands, methods, and performance measures) scheme proposed by \cite{Morris2019}.

\subsection{Aim}
The aim of this simulation is to evaluate ITR estimation for a censored outcome whilst preserving privacy of individual-level data in settings that vary according to whether confounding mechanisms are the same or not across data sites. We therefore focused on the performance of DWSurv parameter estimates, in terms of bias and variability, and the resulting population outcomes under different assumptions concerning: (1) the confounding across sites, (2) the strength of confounding, (3) the censoring mechanism, and (4) model misspecification. Points (1)-(3) concern data-generating mechanisms, while (4) concerns the analytic model selection. Censoring reduces power both due to the loss of observed events and due to some additional variability introduced by inverse probability of censoring weighting. As our focus is on privacy and pooling, most simulation scenarios do not include censoring for simplicity and reduced computational burden; in the scenario that does include censoring, estimators behave as anticipated.

\subsection{Data-generation}
We consider a total of seven data-generated scenarios, with one as our `base' scenario designed to mimic key features of the CPRD, and the remaining scenarios each vary one feature (e.g.~level of confounding, treatment effect, etc.) to explore its impact on results. All scenarios assume 10 sites, with sample sizes accounting for 6-14\% of the total sample size. We consider three covariates, $X_1, X_2, X_3$, and allow their distributions to vary by site. Specifically, we have that
\begin{itemize}
\item $X_1$ follows a Bernoulli(0.55) distribution for odd sites and a Bernoulli(0.4) distribution for even sites,
\item $X_2$ follows a Normal(10,1) distribution for odd sites and a Normal(8,1.5) distribution for even sites,
\item $X_3$ follows a Uniform[6,14] distribution for odd sites and a 7+log-Normal(0.7,0.5) distribution for even sites.
\end{itemize}
Treatment $A$ is binary, and follows a Bernoulli($p(x)$) distribution, where
\[ p\left ( x \right )=[1+e^{-\left ({\alpha_{0,j} + \alpha_{1,j} x_1 + \alpha_{2,j} x_2  + \alpha_{3,j} x_3}  \right )}]^{-1} \]
with $(\alpha_{0,j},...,\alpha_{3,j})$ given in Table \ref{tab:datagen}. 

We consider settings in which the propensity score model is \textit{global} such that all sites' exposure models are generated using the same covariates and the same treatment model parameters (scenarios 1-2); settings in which the propensity score model is \textit{local} such that treatment allocation at all sites depends on all three covariates but the treatment model parameters differ by site (scenarios 3-4); and local settings where some of $(\alpha_{0,j},...,\alpha_{3,j})$ are identically zero such that the confounder-set differs by site (scenarios 5-6). Each scenario differs with its `twin' by the level of the confounding effect. 

The outcome is taken to be a time to serious adverse outcome, $Y$, such that optimizing treatment requires \textit{maximizing} the time to the event. We generate the outcome as
\[	Y = \exp(\beta_0 + \beta_1x_1 + \beta_2 \sin(x_2) + \beta_3x_3 + \beta_4x_1x_3 + a(\psi_0+\psi_1x_2 )+\epsilon),	\]
where $\beta_0 + \beta_1x_1 + \beta_2 \sin(x_2) + \beta_3x_3 + \beta_4x_1x_3$ is the treatment-free function, $a(\psi_0+\psi_1x_2 )$ is the blip function, and the random errors $\epsilon$ follow a standard normal distribution. The parameters $\bm{\beta}$ are given by $(\beta_0,\beta_1,\beta_2,\beta_3,\beta_4) = (4,0.2,-0.1,0.01,-0.005)$ and the blip parameters are $(\psi_0,\psi_1) = (4,-0.55)$ such that the optimal treatment rule is given by `treat with $a=1$ whenever $x_2 < 4/0.55$', or (approximately) `treat with $a=1$ whenever $x_2 < 7.27$', under a large treatment effect setting, or $(\psi_0,\psi_1) = (0.15,-0.015)$ such that the optimal treatment rule is given by `treat with $a=1$ whenever $x_2 < 0.15/0.015$', or (approximately) `treat with $a=1$ whenever $x_2 < 10$', under a small treatment effect setting.

In all scenarios, we assume no censoring. However we consider one final scenario (scenario 7), which corresponds to the base scenario (scenario 5) with censoring at all sites. In this final scenario, the censoring for odd sites occurs randomly with probability 0.3, and at even sites, outcomes are censored with probability $\exp(0.1+0.6X_1)[1+\exp(0.1+0.6X_1)]^{-1}$, leading to approximately 45\% of the sample being censored. 

For each of the seven simulated scenarios, we generated 1000 independent datasets of size $n$ = 2,500 (small sample size) or $n$ = 250,000 (large sample size), further divided in $J=$10 centres, matching the number of centres in the CPRD analysis. All simulations were performed using a customized program in \texttt{R} software that is available in the Online Supplementary Material.


\begin{table}
\caption{Simulations: Data-generating parameters for the propensity score model of the binary treatment $A$. Sc.~denotes `scenario', with 5 the base scenario. All odd-numbered scenarios have moderate confounding, and even-numbered scenarios have strong confounding.} \label{tab:datagen}
\begin{tabular}{lcc}
\hline
 & \multicolumn{2}{c}{\scriptsize{\textbf{Global PS}}} \rule{0pt}{8pt}\\
  & \multicolumn{1}{c}{\scriptsize{\textbf{Sc.~1}}} & \multicolumn{1}{c}{\scriptsize{\textbf{Sc.~2}}} \\
\hline
\scriptsize{$\alpha_{0,j}$} & 0.01 $\forall j$ & 0.1 $\forall j$ \rule{0pt}{12pt}\\
\scriptsize{$\alpha_{1,j}$} & 0.01 $\forall j$ & 0.1 $\forall j$ \rule{0pt}{12pt}\\
\scriptsize{$\alpha_{2,j}$} & 0.01 $\forall j$ & 0.1 $\forall j$ \rule{0pt}{12pt}\\
\scriptsize{$\alpha_{3,j}$} & 0.01 $\forall j$ & 0.1 $\forall j$ \rule{0pt}{12pt}\\
\hline
 & \multicolumn{2}{c}{\scriptsize{\textbf{Local PS, all $X$s}}} \\
  & \multicolumn{1}{c}{\scriptsize{\textbf{Sc.~3}}} & \multicolumn{1}{c}{\scriptsize{\textbf{Sc.~4}}} \\
\hline
\scriptsize{$\alpha_{0,j}$} & U[0.002, 0.02] $\forall j$ & U[0.04, 0.14] $\forall j$\rule{0pt}{12pt}  \\
\scriptsize{$\alpha_{1,j}$} & U[0.002, 0.02] $\forall j$ & U[0.04, 0.14] $\forall j$\rule{0pt}{12pt}  \\
\scriptsize{$\alpha_{2,j}$} & U[0.002, 0.02] $\forall j$ & U[0.04, 0.14] $\forall j$\rule{0pt}{12pt}  \\
\scriptsize{$\alpha_{3,j}$} & U[0.002, 0.02] $\forall j$ & U[0.04, 0.14] $\forall j$\rule{0pt}{12pt}  \\
\hline
 & \multicolumn{2}{c}{\scriptsize{\textbf{Local PS, subset $X$s}}} \\
  & \multicolumn{1}{c}{\scriptsize{\textbf{Sc.~5}}} & \multicolumn{1}{c}{\scriptsize{\textbf{Sc.~6}}}\\
\hline
\scriptsize{$\alpha_{0,j}$} & \multicolumn{1}{c}{U[0.01, 0.06] $\forall j$}  & \multicolumn{1}{c}{U[0.6, 0.8] $\forall j$}  \rule{0pt}{16pt}\\
\scriptsize{$\alpha_{1,j}$}  &   $\left\{
  \begin{array}{ll}
  U[0.01,0.06] \quad & \mbox{ for } j=1,...,4 \\ 
  0 & \mbox{ otherwise }
  \end{array} \right.$  &   $\left\{
  \begin{array}{ll}
  U[0.6,0.8] & \mbox{ for } j=1,...,4 \\ 
  0 & \mbox{ otherwise }
  \end{array} \right.$  \rule{0pt}{16pt}\\
\scriptsize{$\alpha_{2,j}$} &  $\left\{
  \begin{array}{ll}
  U[0.002,0.02] & \mbox{ for } j=4,...,7 \\ 
  0 & \mbox{ otherwise }
  \end{array} \right.$   &
    $\left\{
  \begin{array}{ll}
  U[0.14,0.18] & \mbox{ for } j=4,...,7 \\ 
  0 & \mbox{ otherwise }
  \end{array} \right.$   \rule{0pt}{16pt}\\ 
\scriptsize{$\alpha_{3,j}$} &  $\left\{
  \begin{array}{ll}
  U[0.002,0.02] & \mbox{ for } j=8,9,10 \\ 
  0 & \mbox{ otherwise }
  \end{array} \right.$  &
    $\left\{
  \begin{array}{ll}
  U[0.14,0.18] & \mbox{ for } j=8,9,10 \\ 
  0 & \mbox{ otherwise }
  \end{array} \right.$  \rule{0pt}{16pt}\\
\hline
\end{tabular}

\end{table}


\subsection{Estimands, methods, and performance measures}
The estimands of interest are the blip function parameters $(\psi_0,\psi_1)$, which fully characterize the optimal treatment strategy, and the value function, i.e.~the expected outcome (expected log-survival) in a new population to whom the estimated decision rule is applied.

All analyses relied on DWSurv using overlap weights, but varied in terms of whether or not (i) data aggregation (distributed regression) was performed and (ii) confounding mechanisms differed across sites, that is, whether the propensity score (and censoring models in scenario 7) were estimated globally with all possible confounders, locally with all possible confounders, or locally with known local confounders. Local propensity score and, when needed, censoring models were used in combination with distributed regression, whereas global propensity score and, when needed, censoring models, were used with individual-level data only.

The performance of DWSurv under the various approaches to propensity score and censoring model estimation was assessed assuming one or both of the treatment-free and propensity score models was correctly specified in terms of the included covariates and their functional form. Thus, under a correct treatment-free model, log-time is modelled as a function of $x_1$, $\sin(x_2)$, $x_3$ and the interaction between $x_1$ and $x_3$
\[	\beta_0 + \beta_1x_1 + \beta_2 \sin(x_2) + \beta_3x_3 + \beta_4x_1x_3 + a(\psi_0+\psi_1x_2 )+\epsilon),	\]
whereas the incorrect specification models the log-outcome as a function of linear terms $x_1$, $x_2$, and $x_3$ only but omits the transformation on $x_2$ and the interaction:
\[	\tilde\beta_0 + \tilde\beta_1x_1 + \tilde\beta_2 x_2 + \tilde\beta_3x_3 + a(\psi_0+\psi_1x_2 )+\epsilon).	\]
In the case of correct propensity score specification, the treatment allocation is modelled as a function of either all covariates (global propensity score or local propensity score with all $X$s, respectively referred as ``Global All $X$s'' and ``Local All $X$s'') or only those covariates on which it is known to depend (local propensity score with a subset of $X$s referred as ``Local Some $X$s''). Under incorrect propensity score specification, treatment is assumed to be randomly allocated such that the propensity score model includes only an intercept (referred as ``Global'' or ``Local''). Finally, in the scenario in which censoring occurs, we pair the censoring specification with the propensity score specification, assuming in the incorrect specification setting that censoring occurs completely at random (an intercept only model).

Measures of performance used to assess the estimation strategies are (i) the relative bias of estimators of $(\psi_0,\psi_1)$, (ii) the root mean squared error of the estimators, and (iii) the difference between the value function under the true optimal ITR $\log(Y_{True})$ and estimated optimal ITR $\log(Y_{Opt})$, i.e.~the difference between the expected outcome when the true optimal ITR (derived from data-generating parameters) is applied, relative to the expected outcome under the estimated ITR, and (iv) its empirical standard deviation when the estimated treatment rule was applied to a new cohort of patients of size $n$=100,000.

\subsection{Results}
In this section, we present the results for the seven simulated scenarios with small sample size under various forms of model misspecifications, for a small treatment effect setting and large treatment effect setting. For the sake of space, only results for $\psi_0$, with small treatment effect and small sample size, are presented here; all results relating to $\psi_1$, results for the larger treatment effect settings and results for the larger sample size (with both the moderate and the larger treatment effect settings) are presented in the Online Supplementary Materials.

\begin{landscape}
\begin{figure}
   \begin{center}
        \includegraphics[scale=0.28]{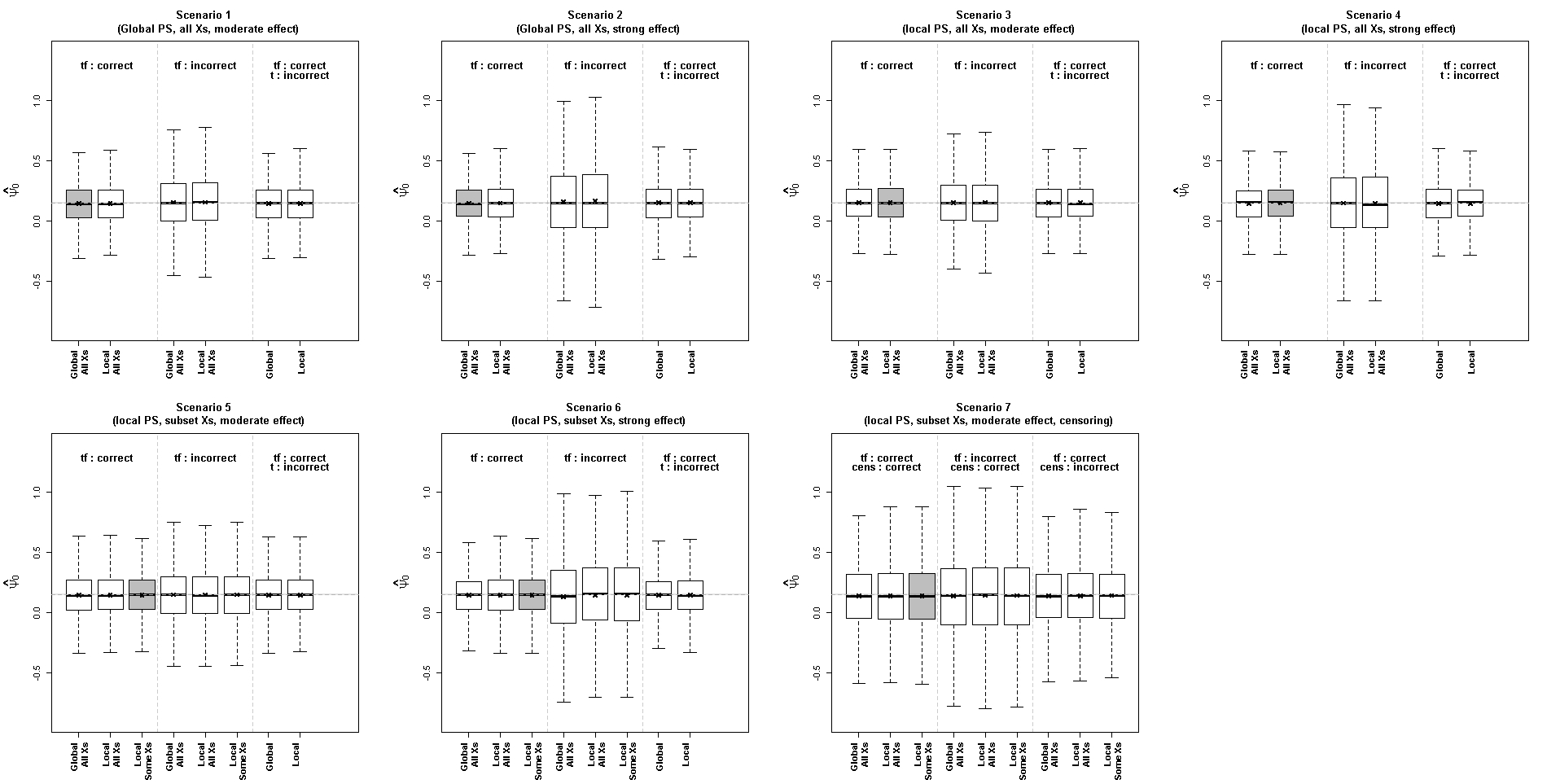}
        \caption{Simulation results: small sample size and small treatment effect settings - Performance of the methods over 1000 iterations in the estimation of $\psi_{0}$. For each scenario, the grey filling indicates the configuration under which the data have been simulated and the light-grey band represents the interval in which the absolute relative bias is less than or equal to 5\%. `PS' denotes propensity score, `tf' denotes treatment-free model, and 'cens' denotes censoring model. An `incorrect' treatment-free model omits the transformation version of $x_2$, $\sin(x_2)$, and the interaction $x_1x_3$ from the specification, while an `incorrect' treatment model assumes random allocation thus fitting an intercept-only propensity score.  }  \label{fig:smallNsmallTx}
    \end{center}
\end{figure}
\end{landscape}

In the smaller sample size setting with a small treatment effect, when the treatment-free is correctly specified, 
all modelling approaches -- global and local -- provide consistent estimation of the blip parameters $(\psi_0,\psi_1)$, with an absolute relative bias less than 5\% and RMSE around 0.03 for $\psi_0$ and around 0.0003 for $\psi_1$ (Table \ref{tab:SmallNLowTmt}, Figure \ref{fig:smallNsmallTx}).

For scenario 7, which corresponds to the base scenario in which a censoring mechanism is included, bias remains low (again, less than 5\%), although higher than for scenario 5, and a higher RMSE is observed (close to 0.07). These results are not surprising, as censoring reduces the effective sample size, and thus, increases variability. Concerning the performance in terms of the value function: when the confounding mechanisms are the same across sites (scenarios 1 and 2) or rely on the same covariates but with different effects across sites (scenarios 3 and 4),  the value function under the estimated optimal treatment is very close to the value function under the true optimal ITR. When the confounding mechanisms differ across sites such that they rely on different covariates in some sites (scenarios 5-7), performance of the ITRs worsens in that the value function under the estimated rule is not as close to the value function under the true optimal ITR, however there is relatively little difference between each of the modelling approaches adopted and the difference is small relative to its standard error (i.e.~with the exception of scenario 7, the difference between the value function under the true and the estimated optimal rule does not significantly differ from 0).


When the treatment-free model is misspecified and the propensity score relies on all or a subset of all possible confounders, under the true-propensity score generation, the methods provide good performance criteria, with a higher variability (Table \ref{tab:SmallNLowTmt}, Figure \ref{fig:smallNsmallTx}). In scenario 2, as the degree of confounding increases compared to scenario 1, absolute relative
bias goes from small to moderate for $\psi_0$ and $\psi_1$; results for $\psi_1$ can be found in the Supplementary Materials. When the confounding mechanisms rely on the same covariates but with different effects across sites (scenarios 3 and 4), the performance of the local and the global propensity score models are equivalent. When the confounding mechanisms do not necessarily rely on the same covariates across sites with high confounding level (scenario 6) and censoring mechanism (scenario 7), the global propensity score method yields a moderate bias as both models (the treatment-free and the propensity score) are not adequate. However, once the propensity score is locally estimated, even with all the possible confounders, the bias is reduced. The base scenario (scenario 5), which differs from scenario 6 and 7 by the strength of the confounding and the censoring respectively, seems to be less sensitive to the misspecifications. Finally, even with the bias and the variability reported in this configuration, the performance in terms of the value function is still satisfactory.

When the treatment-free model is correctly specified and the propensity score is misspecified and estimated under the assumption of random treatment allocation, all methods for all scenarios provide consistent estimation of the blip parameters with performance criteria close to the ones obtained when the treatment-free model is correctly specified. For all configurations, each method provides comparable value functions as reflected by the difference in value function relative to that under the true optimal ITR (Table \ref{tab:SmallNLowTmt}).

In the large treatment effect settings, both local and global approaches to estimation provide consistent estimators of the blip parameters that exhibit less bias than in the small treatment effect setting. 
The differences in value function under the true versus estimated rules are again near zero and similar across methods. See the Online Supplementary Materials for full details.

The effect of increasing the sample size was also investigated (for full results, see the Online Supplementary Materials). These investigations show that the moderate bias that was perceptible in the small sample size setting, especially with a small treatment effect, is no more noticeable and that the variability of the estimates is much lower when the sample size is increased.

The focus of these investigations has been on performance in terms of bias, mean squared error, and the value function. In practice, a measure of variability would also be required. In the Online Supplementary Materials, we provide a formula for a conservative estimate of the standard error (which assumes the weights are fixed rather than estimated), and demonstrates performance relative to the empirical standard deviation of estimates across a limited number of scenarios.

\begin{table}
\caption{Simulation results: Simulations with small sample size and small treatment effect: Performance of the methods over 1000 iterations: mean $\hat{\psi}_0$ and 95\% confidence interval (CI), relative bias (\%, denoted RB) and root mean squared error (RMSE) of $\psi_0$, and difference in value function (dVF) between true and estimated ITR with its standard error (SE). Sc.~denotes scenario. }
\label{tab:SmallNLowTmt}
\begin{tabular}{lllllll}
\hline
{Method} & Sc. & {Mean [CI]} & {RB}  & {RMSE} &  & dVF (SE) \\
\hline
&  & \multicolumn{5}{c}{\scriptsize{\textbf{Treatment-free model correctly specified, propensity score}}}\\
&  & \multicolumn{5}{c}{\scriptsize{\textbf{relies on all or a subset of all possible confounders}}} \\
\hline
Global All $X$s & 1 & 0.144 [-0.197;0.486] & -3.714 & 0.028  &  & -0.006 (0.007)\\
Local All $X$s & 1 & 0.144 [-0.346;0.634] & -4.042 & 0.029  &  & -0.006 (0.007)\\
\hline
Global All $X$s & 2 & 0.148 [-0.170;0.467] & -1.048 & 0.027  & & -0.007 (0.007)\\
Local All $X$s & 2 & 0.149 [-0.548;0.846] & -0.936 & 0.028  & & -0.007 (0.007)\\
\hline
Global All $X$s & 3 & 0.150 [-0.191;0.491] & -0.097 & 0.028  & & -0.006 (0.006)\\
Local All $X$s & 3 & 0.150 [-0.341;0.641] & 0.114 & 0.029  & & -0.006 (0.006)\\
\hline
Global All $X$s & 4 & 0.147 [-0.175;0.469] & -2.127 & 0.026  & & -0.007 (0.007)\\
Local All $X$s & 4 & 0.148 [-0.505;0.801] & -1.428 & 0.027  & & -0.007 (0.007)\\
\hline
Global All $X$s & 5 & 0.150 [-0.193;0.492] & -0.290 & 0.032  & & -0.674 (0.718)\\
Local All $X$s & 5 & 0.149 [-0.341;0.639] & -0.581 & 0.033  & & -0.679 (0.717)\\
Local Some $X$s & 5 & 0.150 [-0.338;0.637] & -0.209 & 0.032  & & -0.674 (0.717)\\
\hline
Global All $X$s & 6 & 0.140 [-0.191;0.488] & -1.250 & 0.028 & & -0.755 (0.739)\\
Local All $X$s & 6 & 0.149 [-0.564;0.862] & -0.577 & 0.033  & &-0.765 (0.737)\\
Local Some $X$s & 6 & 0.150 [-0.560;0.861] & 0.170 & 0.033  & & -0.774 (0.738)\\
\hline
Global  All $X$s & 7 & 0.143 [-0.360;0.646] & -4.690 & 0.067 & & -0.991 (0.076)\\
Local All $X$s & 7 & 0.143 [-0.347;0.634] & -4.570 & 0.073   & & -0.992 (0.080)\\
Local Some $X$s & 7 & 0.144 [-0.344;0.632] & -4.263 & 0.073 & & -0.991 (0.079)\\
\hline
&  & \multicolumn{5}{c}{\scriptsize{\textbf{Treatment-free model misspecified, propensity score relies on  }}} \\
&  & \multicolumn{5}{c}{\scriptsize{\textbf{all or a subset of all possible confounders}}} \\
\hline
Global All $X$s & 1 & 0.155 [-0.285;0.595] & 3.316 & 0.051 & & -0.006 (0.007)\\
Local All $X$s & 1 & 0.154 [-0.476;0.784] & 2.829 & 0.052  & & -0.006 (0.007)\\
\hline
Global All $X$s & 2 & 0.160 [-0.260;0.579] & 6.358 & 0.108 & & -0.008 (0.007)\\
Local All $X$s & 2 & 0.163 [-0.757;1.084] & 8.989 & 0.108  & & -0.008 (0.007)\\
\hline
Global All $X$s & 3 & 0.153 [-0.287;0.592] & 1.881 & 0.051 & & -0.006 (0.007)\\
Local All $X$s & 3 & 0.154 [-0.477;0.785] & 2.745 & 0.052  & & -0.006 (0.007)\\
\hline
Global All $X$s & 4 & 0.147 [-0.277;0.572] & -1.791 & 0.089 & & -0.008 (0.007)\\
Local All $X$s & 4 & 0.148 [-0.711;1.007] & -1.312 & 0.093  & & -0.008 (0.007)\\
\hline
Global All $X$s & 5 & 0.152 [-0.289;0.593] & 1.333 & 0.055 & & -0.726 (0.757)\\
Local All $X$s & 5 & 0.151 [-0.478;0.781] & 0.865 & 0.056  & & -0.725 (0.751)\\
Local Some $X$s & 5 & 0.151 [-0.475;0.778] & 0.911 & 0.055 & & -0.725 (0.756)\\
\hline
Global All $X$s & 6 & 0.134 [-0.302;0.571] & -10.547 & 0.106 & & -0.884 (0.810)\\
Local All $X$s & 6 & 0.150 [-0.776;1.076] & -0.113 & 0.111   & & -0.874 (0.817)\\
Local Some $X$s & 6 & 0.149 [-0.773;1.070] & -0.873 & 0.110  & & -0.873 (0.818)\\
\hline
Global  All $X$s& 7 & 0.140 [-0.495;0.774] & -6.904 & 0.112  & & -0.993 (0.086)\\
Local All $X$s & 7 & 0.144 [-0.486;0.774] & -4.004 & 0.120   & & -0.993 (0.089)\\
Local Some $X$s & 7 & 0.145 [-0.482;0.772] & -3.577 & 0.119  & & -0.992 (0.089)\\
\hline
&  & \multicolumn{5}{c}{\scriptsize{\textbf{Treatment-free model correctly specified, propensity score or }}} \\
&  & \multicolumn{5}{c}{\scriptsize{\textbf{censoring models misspecified (assumed independent of $X$s)}}} \\
\hline
Global  & 1 & 0.145 [-0.199;0.488] & -3.661 & 0.028 & & -0.006 (0.007)\\
Local  & 1 & 0.145 [-0.343;0.632] & -3.540 & 0.028  & & -0.006 (0.007)\\
\hline
Global  & 2 & 0.149 [-0.195;0.494] & -0.587 & 0.029 & & -0.007 (0.007)\\
Local  & 2 & 0.149 [-0.579;0.878] & -0.628 & 0.028  & & -0.007 (0.007)\\
\hline
Global  & 3 & 0.150 [-0.194;0.494] & 0.124 & 0.028 & & -0.006 (0.006)\\
Local  & 3 & 0.150 [-0.338;0.639] & 0.297 & 0.029  & & -0.006 (0.006)\\
\hline
Global  & 4 & 0.146 [-0.199;0.491] & -2.719 & 0.029 & & -0.007 (0.007)\\
Local  & 4 & 0.146 [-0.528;0.820] & -2.575 &  0.028  & & -0.007 (0.007)\\
\hline
Global  & 5 & 0.150 [-0.193;0.492] & -0.097 & 0.032 & & -0.672 (0.718)\\
Local  & 5 & 0.150 [-0.336;0.636] & 0.046 &   0.032   & & -0.674 (0.719)\\
\hline
Global  & 6 & 0.148 [-0.194;0.490] & -1.269 & 0.029 & & -0.753 (0.738)\\
Local  & 6 & 0.148 [-0.578;0.874] & -1.478 & 0.032  & & -0.751 (0.729)\\
\hline
Global & 7 & 0.143 [-0.359;0.645] & -4.506 & 0.067 & & -0.991 (0.076)\\
Local All $X$s  & 7 & 0.143 [-0.347;0.634] & -4.407 & 0.070  & & -0.991 (0.077)\\
Local Some $X$s & 7 & 0.144 [-0.344;0.632] & -4.081 & 0.071  & & -0.990 (0.077)\\
\hline
\end{tabular}

\parbox{18.5cm}{ ``Global All $X$s'' = Global propensity score with all possible confounders}
\parbox{18.5cm}{ ``Local All $X$s'' = Local propensity score with all possible confounders}
\parbox{18.5cm}{ ``Local Some $X$s'' = Local propensity score with known local confounders}
\parbox{18.5cm}{ ``Global'' = Global propensity score with treatment assumed to be randomly allocated}
\parbox{18.5cm}{ ``Local'' = Local propensity score with treatment assumed to be randomly allocated}
\end{table}

\subsection{Observations}
The simulations demonstrate the following key findings: (i) distributed regression analyses better dealt with differences in confounding across sites than global analyses that did not consider local effect and (ii) a high degree of confounding may lead to large weights and more unstable estimations. Not surprisingly, we also observed that the censoring mechanism may induce variability in the estimates, and that performance improves with sample size.

When the treatment-free model is correctly specified, thanks to the double-robustness of DWSurv, all approaches to estimation perform well. However, when the treatment-free is misspecified, the performance varies according to the method. When the true propensity score relies on all possible confounders, all methods provide satisfactory estimation of the blip parameters, although strong confounding can induce some bias in the estimators from the global approach when the true models are in fact local, whereas local estimation can increase variability as seen with the RMSE. When the true propensity score relies on known local confounders, the global approach accounting for all possible confounders yields moderate bias, which is expected, as in this case both treatment-free and treatment models are misspecified such that the conditions needed to ensure double robustness are not satisfied.

Taken together, these results suggest that the distributed regression approach with locally-estimated treatment and censoring models not only provides privacy protection, but it can also lead to better estimation than an analysis relying on full sharing of individual-level data that estimate global models, without site specific covariate effects.

\section{Analysis using the CPRD}
\label{sec:cprd}
We used data from the United Kingdom's CPRD \citep{cprd} to illustrate the different estimation strategies discussed for the optimal individualized treatment rule. Details on the cohort creation, covariates considered, and outcome definition follow.

\subsection{Cohort construction and key variables}
The CPRD is one of the largest primary care databases of de-identified patients data, drawn from a network of more than 700 general practices across the United Kingdom. The database includes data on demographics, lifestyle factors, medical diagnoses (coded via the Read Classification System), referrals to specialists and hospitals, and prescriptions issued by general practitioners. For this analysis, the CPRD data were linked with the Office for National Statistics mortality database as well as the Hospital Episode Statistics (HES) repository, which contains information on diagnoses for each hospital stay, classified according to the International Classification of Diseases, version 10 (ICD-10).  Linkage to HES is restricted to English practices that have consented to the linkage scheme (around 85\%) such that only the corresponding data recorded in England were available for analysis.

We defined a cohort of patients aged 18 years and older, new users of citalopram or fluoxetine between April 1st, 1998 and December 31st, 2017. Cohort entry corresponded to the date of treatment initiation of citalopram or fluoxetine. Patients were required to have at least one year of history in the CPRD prior to cohort entry and an inpatient or outpatient diagnosis of depression in the year before cohort entry. Patients were excluded from the study cohort if they had used any antidepressant drug in the year prior to cohort entry. Patients were followed from cohort entry until an outcome of interest, the administrative end of study (December 31st, 2017), treatment discontinuation for citalopram or fluoxetine, switch to any other antidepressant than the initiating drug or add-on of a second antidepressant, end of registration in the CPRD, or non-suicide death, whichever occurred first.

Our outcome of interest was the first occurrence of a severe depression-related outcome during follow-up, defined as a composite of hospitalisation for depression, hospitalisation for self-harm, or suicide. See \cite{aje} for further details on cohort and variable definitions.  

The `treatment-free' component of the DWSurv model adjusted for the following potential confounders: continuous-valued age in years, sex, an interaction between age and sex, continuous-valued body mass index, smoking, alcohol abuse, calendar time of cohort entry (discretized into 1998-2005, 2006-2011, 2012-2017), a composite indicator of psychiatric disease history (including autism spectrum disorder, obsessive-compulsive disorder, bipolar disorder, schizophrenia, and anxiety or generalized anxiety disorder), separate indicators for previous use of antipsychotics or of benzodiazepines or other psychotropic drugs (which included anxiolytics, barbiturates and hypnotics), prescriptions for lipid-lowering drugs, the number of psychiatric hospital admissions  (including hospitalizations for self-harm) in the 6 months prior to cohort entry, and the quintile of the Index of Multiple Deprivation (\cite{imd}) as a measure of socio-economic status. The smoking status and  body mass index were respectively defined using any most recent smoking status code or drug code for smoking cessation therapy, and any anthropometric data (height, weight, and body mass index when recorded) measured in the 5 years prior to cohort entry. The medication variables were defined using any prescription filled in the year prior. Comorbidities (like psychiatric disease history) were defined using any diagnosis recorded anytime prior to cohort entry. All potential confounders above were also included in the propensity score and censoring models used to compute the overlap and the inverse probability of censoring weights. The same variables (with the exception of the interaction between age and sex) were considered as potential tailoring variables; their possible effect modification was investigated by including interaction terms between each of these variables and the treatment in the outcome model. Missingness in body mass index, smoking status, and the Index of Multiple Deprivation was accounted for by multiple imputation \citep{multipleimp} using fully conditional specification \citep{fullycond} 
 and five imputed datasets.

Data from English CPRD practices are drawn from 10 regions \textit{North East}, \textit{North West}, \textit{Yorkshire and the Humber}, \textit{East Midlands}, \textit{West Midlands}, \textit{East of England}, \textit{South West}, \textit{South Central}, \textit{London}, and \textit{South East Coast}. We considered each of these regions as distinct `centres' in  which prescribing patterns (confounding relationships) and reasons for loss to follow-up (censoring) may differ, such that a distributed regression DWSurv analysis may offer both privacy preservation \textit{and} better control of confounding  and informative censoring.

We described the cohort, proportion of patients from each CPRD English region and, using one imputed dataset, we presented the baseline characteristics of the whole study cohort stratified by 1) the initiating treatment and 2) the censoring indicator. The censoring indicator was equal to 1 if a patient's outcome was never observed. The standardized mean differences (SMDs) were also computed after either weighting patients' data with overlap weights for the treatment or with inverse probability of censoring weights, comparing three strategies for fitting the propensity score and the censoring model.

For the first strategy (termed ``Global''), one propensity score and one censoring model were fitted using the whole study cohort data and were used to estimate the propensity score and censoring probability for each individual in the study. The corresponding weights were included in the DWSurv model and the model fit to the individual level data (using no aggregation). This analysis corresponds to one in which there is no privacy preservation, and data from all centres are shared at an individual record level with a central analysis site.

In the second strategy (termed ``Local''), the propensity score and censoring models were fitted separately in each region using logistic regression models with all potential confounders as predictors in the models, and the treatment rule was fitted using the distributed regression approach to analysis. This strategy allows for privacy preservation, and corresponds to the ``Local All $X$s'' strategy of the simulations.

Finally, the third strategy (termed ``Local with selection'') was based on a regional selection of confounders and censoring predictors for the weights only, and used the distributed regression approach to analysis, in which the DWSurv treatment-free model included all potential confounders. In the Online Supplementary Materials, we present in heat maps the variables selected for the weights in each region and their frequency of selection across all 5 imputed datasets (since selection could vary across imputations). To avoid including instrumental variables in the treatment models, variables were selected if they were statistically significant (p-value$<$0.05) in both respective logistic or Cox univariate models predicting the treatment and the outcome to be included in a regional treatment model. Similarly, to be included in a regional censoring model, variables had to be statistically significant in both respective logistic and Cox univariate models for the censoring and the outcome fitted to that regional data. Variables that met those criteria were included in the respective regional (local) treatment or censoring models and separate propensity scores and censoring probabilities were fitted. The third strategy allows each center (region) to work independently as they can fit their own propensity score and censoring model without relying on a global model or a global selection of variables to be included as predictors in the models. This analysis is similar to the ``Local Some $X$s'' of the simulation, but of course the true confounders are not known in this real-data analysis.

For all three analytic strategies, the coefficients in the blip functions were averaged across all 5 analyses corresponding to the 5 imputed datasets, to provide three final estimated individualized treatment rules. The individualized rules were then compared in terms of their average coefficients across all five imputed datasets (for each variable in the rule), and after combining (pooling) all data across all 5 imputations in a whole dataset, the percent concordant in the optimal estimated treatment decisions,  and the median predicted time to SDO under each rule's optimal treatment decision.

\subsection{Results}
The final cohort comprised 246,503 patients of whom 137,791 (53\%) initiated citalopram. A flow chart of the cohort creation can be found in \cite{aje}.  The proportion of patients from each region varied from 2\% in \textit{North East} up to 16\% in \textit{North West} (see Online Supplementary Materials).  In general, patients initiating citalopram were similar to those initiating fluoxetine, with some differences in age, calendar year at cohort entry, proportion of patients with a diagnosed psychiatric disease, and use of lipid lowering drugs (top of Table \ref{tab:char}). As expected, the SMDs were reduced after using treatment overlap weights from any of the 3 estimation strategies, but some differences remained across the groups (SMDs$>$0.10) in the calendar year and the indicator of psychiatric disease (top of Table \ref{tab:char}).

In the cohort stratified by censoring indicator, we found important differences in age, sex, Index of Multiple Deprivation, calendar year, body mass index, smoking status, alcohol abuse, proportion of patients with a diagnosed psychiatric disease, and the number of psychiatric admissions (all corresponding SMDs$>$0.10, bottom of  Table \ref{tab:char}). Weighting the data with the different versions of the inverse probability of censoring weights mostly improved the balance of the variables age, Index of Multiple Deprivation, calendar year, and alcohol abuse, with other variables' balance remaining unchanged. While this lack of improvement is some cause for concern, we anticipate that the treatment-free model may help to alleviate residual bias due to these imbalances. While it is possible that our censoring model is functionally too simple, another plausible explanation is that the proportion of the sample that is censored is so high that the censoring model is dominated by its intercept.

\begin{figure}
\begin{center}
\includegraphics[width=12cm]{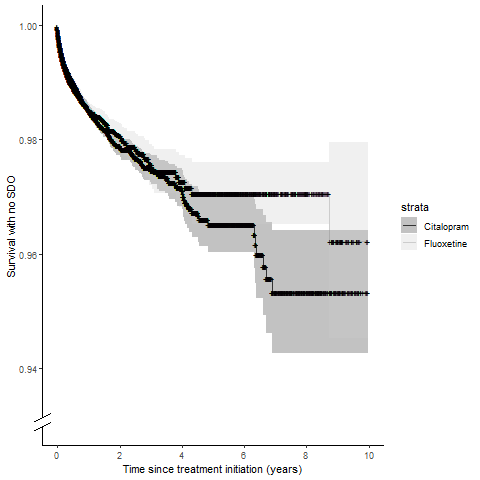}
\end{center}
\caption{Unadjusted Kaplan-Meier curves for the time to SDO over the first 10 years of follow-up stratified by initiating treatment, Clinical Practice Research Datalink, United Kingdom, 1998-2017} \label{kmfig}
\end{figure}

The median follow-up time was 103 days in citalopram users and 93 days in fluoxetine users. When restricting the cohort to only the patients who had an observed SDO, the medians dropped to 45 days and 47 days, respectively. That is, the median time to an event (rather than to end of follow-up) was approximately 6.5-7 weeks. In citalopram users, 1371 SDOs (26\% depression, 71\% self-harm, 4\% suicide) occurred over 80,907 person-years (PYs) of follow-up, yielding an SDO rate of 1.69 per 100PY. Fluoxetine users presented with a similar rate of 1.67 per 100PY, with 920 SDOs (23\% depression, 74\% self-harm, 4\% suicide) occurring over 54,781 PY. Figure \ref{kmfig} presents the unadjusted Kaplan-Meier curves for the time to SDO stratified by the initiating treatment.

\begin{table}
\caption{Baseline characteristics from one imputed dataset stratified by initiating treatment (top) or censoring indicator (bottom) and the corresponding standardized mean differences computed before and after re-weighting by the different overlap weights or inverse probability of censoring weights to account for covariate imbalances between treatment or censoring groups, Clinical Practice Research Datalink, United Kingdom, 1998-2017}
\label{tab:char}
\begin{tabular}{lcccccc}
\hline\noalign{\smallskip}
  & \multicolumn{2}{c}{Treatment} &  & &   &  \\
Variable$^a$ & Citalopram & Fluoxetine & SMD$^c$  & SMD$^{d}$ & SMD$^{e}$ & SMD$^{f}$ \\ \hline
 Age$^b$& 43.4 (18.2)& 40.7 (16.3)& 0.15 & 0.03 & 0.02 & 0.07   \\
 Female& 87,561 (64) &71,826 (66) & 0.05& 0.05 &0.05 &0.04   \\
Age $\times$  female$^b$  &27.3 (25.5) & 26.3 (23.1) & 0.04& 0.01&0.02 & 0.01  \\
Index of Multiple Deprivation$^b$  & 3.0 (1.4)& 3.1 (1.4)& 0.03&0.02 &0.01 &0.13   \\
Calendar year   && & 0.50 & 0.51 & 0.50& 0.49  \\
\hspace{0.2cm} 1998-2005    & 40,077 (29)& 56,502 (52)& & & &  \\
\hspace{0.2cm} 2006-2011   &65,297 (47)& 38,780 (36)& & & & \\
\hspace{0.2cm} 2012-2017 &32,417 (24)& 13,430 (12) &&&&  \\
Body mass index$^b$ &26.8 (5.7)&26.7 (5.7)&0.01&0.00&0.00&0.00 \\
Non-smoker    &55,090 (40) &42,609 (39) & 0.02 &0.01  &0.01 &0.02   \\
 Alcohol abuse  &11,008 (8)&7130 (7) &0.06 & 0.06&0.06 & 0.02  \\
Psychiatric disease   &43,995 (32)&25,501 (24) &0.19 & 0.19&0.18 &0.15   \\
 Medication & & & & & & \\
\hspace{0.2cm} Antipsychotics use   &16,078 (12)&11,499 (11) &0.04 &0.01 & 0.01 &   0.00 \\
 \hspace{0.2cm} Benzodiazepines use  & 27,169 (20)&18,579 (17) &0.07 & 0.04& 0.04 &   0.05 \\
 \hspace{0.2cm} Lipid lowering drugs use &10,487 (8) &5385 (5)& 0.11& 0.07&0.07 &0.06   \\
 No.~psychiatric admissions$^b$ & 0.1 (0.3)& 0.0 (0.5) & 0.04& 0.03&  0.03 &  0.03  \\
      \hline
  & \multicolumn{2}{c}{Censoring} &  & &   &  \\
  & Yes & No & SMD$^c$  & SMD$^{g}$ & SMD$^{h}$ & SMD$^{i}$ \\ \hline
 Age$^b$  & 42.2 (17.5)& 39.0 (17.5) &0.19& 0.09&0.11 & 0.06  \\
 Female & 158,079 (65) & 1308 (57) &0.16& 0.17& 0.17 &   0.23 \\
Age $\times$  female$^b$ & 26.9 (24.5)& 21.7 (23.2)& 0.22& 0.10 &0.09 &  0.15  \\
 Index of Multiple Deprivation$^b$    & 3.0 (1.4)& 3.3 (1.4)& 0.22& 0.18  &  0.19 & 0.10  \\
 Calendar year   && &0.16& 0.13 & 0.12 &  0.11  \\
\hspace{0.2cm} 1998-2005     &95,849 (39)& 730 (32) && & &   \\
\hspace{0.2cm} 2006-2011    &102,957 (42) &1120 (49) && & &   \\
\hspace{0.2cm} 2012-2017 & 45,406 (19)&441 (19) && & &   \\
 Body mass index$^b$ & 26.7 (5.7)&26.1 (5.6) &0.12&0.10  &0.10 &  0.10  \\
Non-smoker & 96,969 (40) &730 (32) &0.16&  0.14& 0.14 &   0.16 \\
 Alcohol abuse   &17,559 (7) &579 (25) &0.51&  0.49& 0.48 &   0.45 \\
     Psychiatric disease   & 68,587 (28)&909 (40) &0.25&  0.26 & 0.28 &  0.24  \\
     Medication & & & & & & \\
  \hspace{0.2cm} Antipsychotics use   & 27,314 (11)& 263 (11) &0.01& 0.05& 0.05 & 0.05   \\
 \hspace{0.2cm} Benzodiazepines use   & 45,316 (19)  &432 (19) &0.01& 0.06 & 0.06 &  0.06  \\
  \hspace{0.2cm} Lipid lowering drugs use  & 15,744 (6)& 128 (6) &0.04&  0.06& 0.07 & 0.05   \\
 No.~psychiatric admissions$^b$  &0.04 (0.36) & 0.33 (1.04)&0.38&0.39  &0.42 &   0.39 \\
     \hline
 \end{tabular}

 \scriptsize{
 \vspace{0.01cm}
 a. Frequencies (\%) presented, unless otherwise stated. \\
 b. Mean (SD).  \\
 c. Computed on the original imputed dataset. \vspace{0.07cm}\\
Computed on the imputed dataset re-weighted by:\\
 d. an overlap weight based on a global propensity score.\\
 e. an overlap weight based on a local propensity score.\\
 f. an overlap weight based on a local propensity score after local variable selection.  \\
 g. an inverse probability of censoring weight based on a global censoring model.\\
 h. an inverse probability of censoring weight based on a local censoring model.\\
 i. an inverse probability of censoring weight based on a local censoring model after local variable selection.}
\end{table}

\begin{figure}
\begin{center}
\includegraphics[width=15cm]{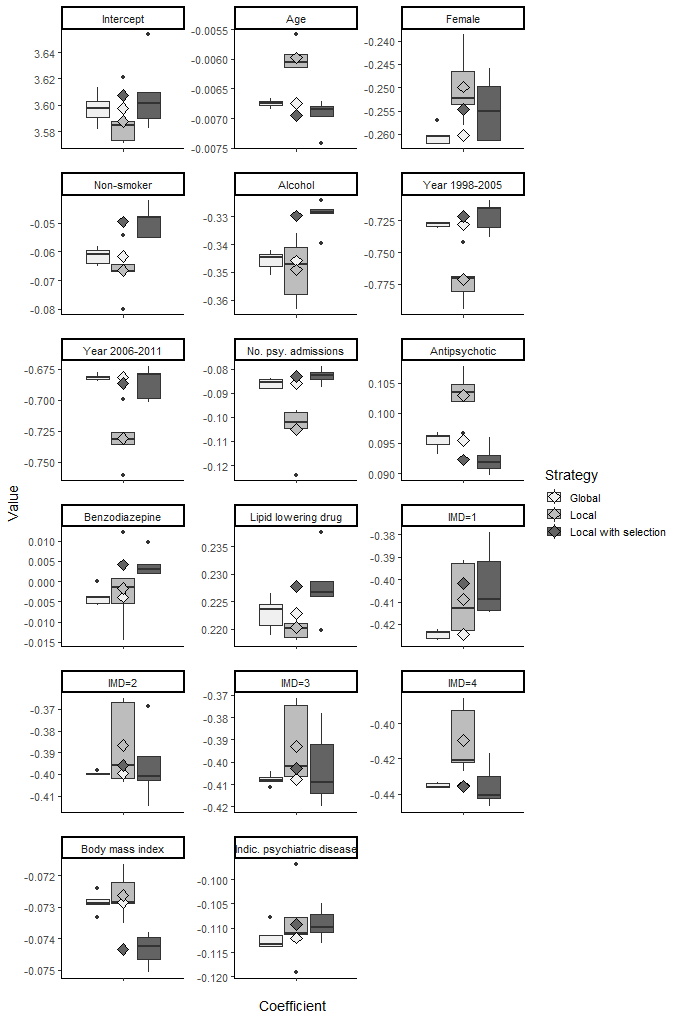}
\end{center}
\caption{Comparison of the coefficients in the blip functions across all 3 strategies for fitting the treatment and the censoring models. Each boxplot summarizes 5 estimated coefficients corresponding to 5 imputed datasets. The mean across all 5 imputed datasets and for a given strategy is depicted by a diamond of the corresponding color in the central boxplot, Clinical Practice Research Datalink, United Kingdom, 1998-2017} \label{figcoef}
\end{figure}

Coefficients in the individualized treatment rules did not vary much across the imputed datasets for a given estimation strategy (Figure \ref{figcoef}) or across the three estimation strategies for the individualized treatment rule (Table \ref{tabcoef}, see the average coefficients corresponding to three final individualized treatment rules): although the boxplots in some cases appear to be quite different, the scale of the plots are in all cases quite small.  We also observed similar results when comparing the predicted median time to SDO under the different individualized treatment rules. In patients with an observed SDO, using the optimal treatment decision led to a predicted gain of 4-6 days in the time to SDO, as opposed to the actual treatment received (Table \ref{tabtimeto}). Furthermore, the concordance in optimal treatment decisions was high for any pair of estimation strategies we compared, with 82\% concordance for the strategies ``Global'' and ``Local'', 84\% for ``Global'' and ``Local with selection'', and 80\% for ``Local'' and ``Local with selection'' strategies.

In this study, although there were some differences in the confounding and censoring mechanisms across the CPRD English regions (Online Supplementary Material), using aggregated data and regional models with or without variable selection, or a more global approach with individual patient data and common censoring and treatment models to develop the treatment rule led to comparable results. Using the aggregated data and local nuisance models therefore offers a means of preserving patients' privacy across centers while allowing for the possibility of variations in confounding and informative censoring; the size of our sample ensures that any reduction in bias due to a local, selected-variable approach is not compromised by any notable loss of efficiency.

\begin{table}
\caption{Average coefficients from the blip function across 5 imputations and for each estimation strategy for the treatment and censoring models, Clinical Practice Research Datalink, United Kingdom, 1998-2017}
\label{tabcoef}
\begin{tabular}{lccc}
\hline\noalign{\smallskip}
Coefficient&Global models  & Local models & Local models and selection  \\ \hline
Intercept (citalopram) & 2.77&2.49&2.50\\
Age &-0.01 &-0.01 &-0.00  \\
Female &-0.37 &-0.28 &-0.36  \\
Index of Multiple Deprivation & & &  \\
\hspace{0.5cm}1 & -0.28 & -0.17& -0.26 \\
\hspace{0.5cm}2 & -0.21 & -0.29 & -0.12  \\
\hspace{0.5cm}3 & -0.17& -0.26 & -0.20  \\
\hspace{0.5cm}4 & -0.21& -0.36 &-0.09  \\
Calendar year & & &  \\
\hspace{0.5cm}1998-2005 &-0.58 & -0.67  & -0.52  \\
\hspace{0.5cm}2006-2011 & -0.41 & -0.35 & -0.28 \\
Body mass index & -0.05 &-0.05 &-0.05  \\
Non-smoker & -0.18&-0.05 &-0.16  \\
Alcohol abuse & -0.24&-0.09 &-0.17  \\
 Indicator psychiatric disease & -0.30& -0.29 & -0.41  \\
Medication & &  &  \\
 \hspace{0.5cm}Antipsychotics use &0.07 & -0.05 &0.19  \\
\hspace{0.5cm}Benzodiazepines use& 0.13 & 0.27 & 0.21  \\
\hspace{0.5cm}Lipid lowering drugs use& 0.36 &0.03 & -0.23  \\
 Number psychiatric admissions &-0.02 &-0.05 &-0.07  \\
\noalign{\smallskip}\hline
\end{tabular}
\end{table}

\begin{table}
\caption{Predicted median time to SDO in days (interquartile range) under different treatment scenarios and for each estimation strategy, Clinical Practice Research Datalink, United Kingdom, 1998-2017}
\label{tabtimeto}
\begin{tabular}{lccc}
\hline\noalign{\smallskip}
  &Global    & Local  & Local models    \\
Scenario &  models  &   models&  and selection  \\ \hline
Everyone treated with the treatment they received&  43 (37-50) & 43 (35-51)   & 42 (35-50)  \\
Everyone treated with citalopram & 42 (35-51) &40 (32-51) & 41 (34-51) \\
Everyone treated with fluoxetine &  44 (38-50)&  45 (39-51)&  43 (37-50)\\
Everyone treated with the optimal treatment$^a$ & 47 (41-54) &48 (41-56)   & 48 (41-56)\\
\noalign{\smallskip}\hline
\end{tabular}

\scriptsize{Abbreviation: SDO, severe depression-related outcome. \\
a. According to the estimated dynamic treatment regime.}
\end{table}

\section{Discussion} \label{sec:discuss}
In order to provide patients the evidence-based medicine, defined as ``judicious use of current best evidence in making decisions about the care of individual patients,''
the development of ITRs must be undertaken using high quality data and appropriate analytic methods. As ITRs seek to detect treatment-covariate interactions, large datasets are required, particularly when the outcome may be censored such that power is a function not of the sample size, but rather of the number of observed events. This may require collaboration across different data centres or even collaboration between countries. Considering both the sensitive and valuable nature of data, ensuring privacy of individual-level data may make collaboration or data-sharing more enticing. We evaluated the performance of distributed regression, adapted to estimate optimal individualized treatment rules when the outcome is a survival time subject to right-censoring and when confounding and censoring mechanisms may be different across data sites. Of course, it is possible to adopt a local confounding approach with the sharing of individual data, either through interactions between site and covariates or simply performing a `local' modelling analysis in the central analysis site. However, while this approach offers the same confounding control as the distributed regression approach, it does not offer the same data protection. We found that in large samples, the distributed regression approach ensures data privacy and good (local) control of confounding without serious loss of precision -- very nearly a proverbial free lunch.

We note that in our simulations, we considered only scenarios where `direction' of confounding was the same across all data sites, even if the models differed, such that all propensity score parameters were non-negative. While this setting is plausible, there may be situations in which confounding relationships are more strikingly different, with a given covariate increasing the probability of receiving treatment in some centres, and decreasing the probability in others. In these settings, it is likely that the advantage of a local modelling approach over a global approach would be more striking than what we have observed.

Our case study using the CPRD demonstrates an approach that can be used to estimate ITRs for mental health care using large, aggregated data sources without compromising patient confidentiality or control of confounding or potentially covariate-driven censoring. Unfortunately, our analysis should be viewed as a case study highlighting methodological contributions rather than in light of its clinical findings, as there remains a strong possibility of unmeasured confounding. In particular, information on the severity of depression is not available in the CPRD, and so we could only adjust for surrogates of this using the number of psychiatric admissions or hospitalisations for self-harm in the six months preceding cohort entry. Second, we had no information on race or ethnicity, which could be associated with both antidepressant treatment prescribed, and outcomes \citep{ethnic1,ethnic2}. Finally, it is likely that information on concomitant psychotherapy was not fully recorded and therefore, that information was not used in our analysis. Methods for sensitivity analyses to assess the impact of violations of the no unmeasured confounding assumption are currently under development.

Another significant challenge in our analysis of the CPRD was that we were unable to model the censoring mechanism well enough to remove all imbalances in the covariates between those individuals who were and were not censored. It is possible that a more flexible model with interactions may have improved performance, and a less parametric model may have improved balance. This must be balanced against the possibility of estimating a model that has superior predictive power, analogous to the setting where a complex propensity score with high predictive power may prioritize instruments over confounding variables and thereby fail to improve over simpler, coarser models when it comes to covariate balance \citep{AlamPS2019}.

In our ``local'' analysis of the CPRD, each region's confounding variables were chosen rather naively, by selecting those covariates that were statistically significantly associated with both the treatment and the outcome in univariate models; covariates included in the censoring model were similarly selected. While the sample size of the CPRD is such that even quite small effects may be deemed significant, this approach would not in general be recommended. Recent proposals such as the outcome-adaptive lasso method \citep{ShortreedAOL2017} of selecting confounding variables would be a better alternative provided each centre had sufficient expertise to implement a more sophisticated approach such as this. It would be important to understand the local statistical resources and skills to determine what methods to use if a local  modelling approach to confounding is adopted. It may also be of interest to study the impact of using different approaches to confounder selection at different centres on the estimated treatment rule.

An important consideration of our statistical approach lies in the estimation of standard errors. Under the distributed regression approach, standard errors can be computed with relatively little difficulty under a conservative approach that ignores the estimation of parameters in the two nuisances models, i.e.~without accounting for estimation of the treatment and censoring models, using a meta-analytic approach. This is an approach that is frequently employed in, for example, marginal structural models and is unlikely to have substantial impact in the very large datasets that are likely to be analyzed using a multi-centre, distributed regression approach, although the impact of censoring has yet to be carefully explored. It would be of interest to examine the feasibility of alternative approaches such as the bootstrap approach of \citet{ShuPrivacy2020}, which requires the sharing of multiple bootstrapped data aggregated matrices and may be computationally infeasible. However, we note that a bootstrap approach requires resampling and sharing of multiple aggregate datasets from each data centre, which is not very appealing from a practically standpoint, even if the statistical properties of the approach are good.

The distributed regression approach assumes the relationship between these variables and the outcome is the same across centres -- further implying that any tailoring effects are common across centres. This is a plausible assumption, as it implies that the (possibly biological) mechanisms that govern the outcome distribution are stable or fixed by nature. Thus, while the `case-mix' (covariate distribution) may vary from site to site, and the treatment allocation procedures (essentially, the propensity score) may also depend on local practice, it is reasonable to posit that the impact of treatment and individual-level factors on the outcome do not vary by geography or health system specific practices.

Distributed regression also requires the use of the same variables in the treatment-free and blip model design matrices at each centre. This requirement may be difficult to satisfy if the centres contributing data are not from a common network, e.g., centres across the world contributing to a cancer or transplant registry rather than different hospitals within the same health maintenance organization. In practice, this limits the design matrix for the treatment-free model to be restricted to include only the predictors that are common to all centres. However, this does not restrict the confounding adjustment to rely on common variables since the propensity scores and weighting matrices are computed locally. Therefore, assuming each centre has measured all relevant centre-specific confounders to account for local treatment allocation patterns, restricting the treatment-free model to common variables should not affect consistency thanks to the double robustness of DWSurv although this could reduce efficiency.

An interesting avenue for future research would investigate the use of distributed regression to estimate \textit{partially} adaptive treatment strategies \citep{TalbotPATS2022}. These are treatment rules in which some known tailoring variables are deliberately omitted from the individualized treatment rule -- a situation which could arise if, say, a particular covariate or biomarker were expensive or frequently unavailable to prescribers. This approach could also be useful if centres collect different covariates, so that tailoring on a common set of covariates is not possible.

DWSurv was developed to model the impact of treatment (and covariates) on event times in an uncensored population, and therefore adjusts only for the \textit{fact} of censoring (binary); an interesting avenue of future research would be to explore modeling the \textit{time to} censoring. A further and very important venue for future work is to establish whether distributed regression can be used to model adaptive treatment strategies - that is sequences of ITRs that account for evolving patient characteristics.

The quest for tailored treatment strategies to improve patient outcomes and, in some cases, reduce treatments costs requires large datasets to detect what are very often small heterogeneities in treatment effects. This may require combining data from different sources such as different sites, study types (e.g.~randomized and observational), or perhaps even healthcare systems. When sites are reluctant or unable to share individual patient data, or when confounding relationships differ across data sources, distributed regression combined with DWSurv can be used to adjust for confounding and covariate-dependent censoring while ensuring data privacy to produce unbiased estimators of individualized treatment rules of censored outcomes.

\small 
\newpage 

{\large Supplementary Material : Privacy-preserving estimation of an optimal individualized treatment rule : A case study in maximizing time to severe depression-related outcomes}

\section*{A. R code to reproduce the simulation study}
Please contact authors Danieli or Moodie.

\section*{B. Additional simulation results}
In what follows, we provide additional tables of results for both parameters of the optimal individualized treatment rule and the resulting value function in the seven scenarios described in the main article for small and large sample sizes, and small and large treatment effects.


\scriptsize
\begin{table}\label{tab:SmallNLowTmt}
{
\caption{Simulation results: Simulations with small sample size and small treatment effect: Performance of the methods over 1000 iterations: mean estimate (95\%CI), relative bias (\%, denoted RB) and root mean squared error (RMSE) of $\hat\psi_1$, and difference in value function (dVF) between true and estimated ITR with its standard error (SE). Sc.~denotes scenario. }
\begin{tabular}{llllllll}
\hline
{Method} & Sc. & && & {Mean [CI]} & {RB}  & {RMSE}  \\
\hline
&  &  \multicolumn{6}{c}{\scriptsize{\textbf{Treatment-free model correctly specified, propensity score}}}\\
&  &  \multicolumn{6}{c}{\scriptsize{\textbf{relies on all or a subset of all possible confounders}}}

\\
\hline
Global All $X$s & 1 & &&&-0.014 [-0.051;0.023] & -4.350 & $3.30\times10^{-4}$  \rule{0pt}{8pt}\\
Local All $X$s & 1 & & &&-0.014 [-0.068;0.039] & -4.527 & $3.30\times10^{-4}$  \rule{0pt}{8pt}\\

\hline

Global All $X$s & 2 && && -0.015 [-0.050;0.021] & -2.193 & $2.90\times10^{-4}$\rule{0pt}{8pt}\\
Local All $X$s & 2 & && &-0.015 [-0.092;0.063] & -2.271 & $3.10\times10^{-4}$  \rule{0pt}{8pt}\\

\hline

Global All $X$s & 3 &&& & -0.015 [-0.052;0.022] & -0.081 & $3.40\times10^{-4}$ \rule{0pt}{8pt}\\
Local All $X$s & 3 &&& & -0.015 [-0.068;0.038] & 0.148 & $3.50\times10^{-4}$ \rule{0pt}{8pt}\\

\hline

Global All $X$s & 4 &&&& -0.015 [-0.050;0.021] & -0.947 & $2.90\times10^{-4}$  \rule{0pt}{8pt}\\
Local All $X$s & 4 && && -0.015 [-0.087;0.057] & -0.033 & $3.00\times10^{-4}$  \rule{0pt}{8pt}\\

\hline

Global All $X$s & 5 &&& & -0.015 [-0.052;0.022] & -0.367 & $3.60\times10^{-4}$  \rule{0pt}{8pt}\\
Local All $X$s & 5 &&& & -0.015 [-0.068;0.038] & -0.815 & $3.70\times10^{-4}$  \rule{0pt}{8pt}\\
Local Some $X$s & 5 && && -0.015 [-0.068;0.038] & -0.390 & $3.60\times10^{-4}$  \rule{0pt}{8pt}\\

\hline

Global All $X$s & 6 && && -0.014 [-0.051;0.022] & -3.340 & $3.00\times10^{-4}$  \rule{0pt}{8pt}\\
Local All $X$s & 6 & &&& -0.015 [-0.092;0.063] & -2.150 & $3.70\times10^{-4}$  \rule{0pt}{8pt}\\
Local Some $X$s & 6 &&& & -0.015 [-0.092;0.063] & -1.276 & $3.60\times10^{-4}$  \rule{0pt}{8pt}\\

\hline

Global& 7 && && -0.014 [-0.067;0.039] & -6.105 & $7.40\times10^{-4}$ \rule{0pt}{8pt}\\
Local All $X$s  & 7 && && -0.014 [-0.067;0.039] & -5.869 & $8.10\times10^{-4}$  \rule{0pt}{8pt}\\
Local Some $X$s  & 7 &&& & -0.014 [-0.067;0.039] & -5.747 & $8.10\times10^{-4}$  \rule{0pt}{8pt}\\

\hline
&  & \multicolumn{6}{c}{\scriptsize{\textbf{Treatment-free model misspecified, propensity score relies on  }}} \\
&  & \multicolumn{6}{c}{\scriptsize{\textbf{all or a subset of all possible confounders}}} \\
\hline
Global All $X$s & 1 & &&& -0.016 [-0.064;0.033] & 3.449 & $6.10\times10^{-4}$ \rule{0pt}{8pt}\\
Local All $X$s & 1 && && -0.015 [-0.085;0.054] & 3.063 & $6.30\times10^{-4}$ \rule{0pt}{8pt}\\

\hline

Global All $X$s & 2 & &&& -0.016 [-0.063;0.031] & 6.259 & $1.37\times10^{-3}$ \rule{0pt}{8pt}\\
Local All $X$s & 2 & &&& -0.016 [-0.120;0.087] & 8.910 & $1.37\times10^{-3}$ \rule{0pt}{8pt}\\

\hline

Global All $X$s & 3 &&&&  -0.015 [-0.064;0.033] & 2.037 & $6.20\times10^{-4}$ \rule{0pt}{8pt}\\
Local All $X$s & 3 && && -0.015 [-0.085;0.054] & 3.058 & $6.30\times10^{-4}$ \rule{0pt}{8pt}\\

\hline

Global All $X$s & 4 & &&& -0.015 [-0.062;0.033] & -1.071 & $1.12\times10^{-3}$ \rule{0pt}{8pt}\\
Local All $X$s & 4 && && -0.015 [-0.111;0.081] & -0.266 & $1.17\times10^{-3}$ \rule{0pt}{8pt}\\

\hline

Global All $X$s & 5 && && -0.015 [-0.064;0.033] & 1.341 & $6.50\times10^{-4}$ \rule{0pt}{8pt}\\
Local All $X$s & 5 && && -0.015 [-0.084;0.054] & 0.712 & $6.50\times10^{-4}$ \rule{0pt}{8pt}\\
Local Some $X$s & 5 && && -0.015 [-0.084;0.054] & 0.781 & $6.50\times10^{-4}$\rule{0pt}{8pt}\\

\hline

Global All $X$s & 6 & &&& -0.013 [-0.061;0.035] & -12.916 & $1.29\times10^{-3}$ \rule{0pt}{8pt}\\
Local All $X$s & 6 & &&& -0.015 [-0.117;0.087] & -1.889 & $1.35\times10^{-3}$ \rule{0pt}{8pt}\\
Local Some $X$s & 6 & &&& -0.015 [-0.116;0.087] & -2.596 & $1.33\times10^{-3}$ \rule{0pt}{8pt}\\

\hline

Global& 7 && && -0.014 [-0.082;0.054] & -8.448 & $1.26\times10^{-3}$ \rule{0pt}{8pt}\\
Local All $X$s  & 7 & &&& -0.014 [-0.083;0.055] & -5.170 & $1.36\times10^{-3}$ \rule{0pt}{8pt}\\
Local Some $X$s  & 7 & &&& -0.014 [-0.083;0.055] & -4.895 & $1.35\times10^{-3}$ \rule{0pt}{8pt}\\

\hline
&  & \multicolumn{6}{c}{\scriptsize{\textbf{Treatment-free model correctly specified, propensity score or }}} \\
&  & \multicolumn{6}{c}{\scriptsize{\textbf{censoring models misspecified (assumed independent of $X$s)}}} \\
\hline
Global  & 1 & &&& -0.014 [-0.052;0.023] & -4.285 & $3.30\times10^{-4}$ \rule{0pt}{8pt}\\
Local  & 1 & &&& -0.014 [-0.067;0.039] & -4.118 & $3.30\times10^{-4}$ \rule{0pt}{8pt}\\

\hline

Global  & 2 & &&& -0.015 [-0.052;0.023] & -1.774 & $3.10\times10^{-4}$ \rule{0pt}{8pt}\\
Local  & 2 & &&&-0.015 [-0.095;0.065] & -1.898 & $3.10\times10^{-4}$ \rule{0pt}{8pt}\\

\hline

Global  & 3 & &&& -0.015 [-0.052;0.022] & 0.151 & $3.40\times10^{-4}$ \rule{0pt}{8pt}\\
Local  & 3 &  &&&-0.015 [-0.068;0.038] & 0.274 & $3.40\times10^{-4}$ \rule{0pt}{8pt}\\

\hline

Global  & 4 && && -0.015 [-0.052;0.023] & -1.637 & $3.20\times10^{-4}$ \rule{0pt}{8pt}\\
Local  & 4 && && -0.015 [-0.089;0.059] & -1.551 & $3.20\times10^{-4}$ \rule{0pt}{8pt}\\

\hline

Global  & 5 &&& & -0.015 [-0.052;0.022] & -0.167 & $3.60\times10^{-4}$ \rule{0pt}{8pt}\\
Local  & 5 & &&& -0.015 [-0.068;0.038] & -0.054 & $3.60\times10^{-4}$ \rule{0pt}{8pt}\\

\hline

Global  & 6 && && -0.014 [-0.052;0.023] & -3.380 & $3.10\times10^{-4}$ \rule{0pt}{8pt}\\
Local  & 6 &&&& -0.015 [-0.093;0.064] & -3.251 & $3.60\times10^{-4}$ \rule{0pt}{8pt}\\

\hline

Global& 7 & &&& -0.014 [-0.067;0.039] & -5.930 & $7.40\times10^{-4}$ \rule{0pt}{8pt}\\
Local All $X$s  & 7 & &&& -0.014 [-0.067;0.039] & -5.749 & $7.90\times10^{-4}$ \rule{0pt}{8pt}\\
Local Some $X$s  & 7 & &&&-0.014 [-0.067;0.039] & -5.570 & $7.90\times10^{-4}$ \rule{0pt}{8pt}\\

\hline
\end{tabular}
}

\parbox{18.5cm}{ ``Global All $X$s'' = Global propensity score with all possible confounders}
\parbox{18.5cm}{ ``Local All $X$s'' = Local propensity score with all possible confounders}
\parbox{18.5cm}{ ``Local Some $X$s'' = Local propensity score with known local confounders}
\parbox{18.5cm}{ ``Global'' = Global propensity score with treatment assumed to be randomly allocated}
\parbox{18.5cm}{ ``Local'' = Local propensity score with treatment assumed to be randomly allocated}
\end{table}

\small 
\begin{landscape}
\begin{figure}
   \begin{center}
        \includegraphics[scale=0.28]{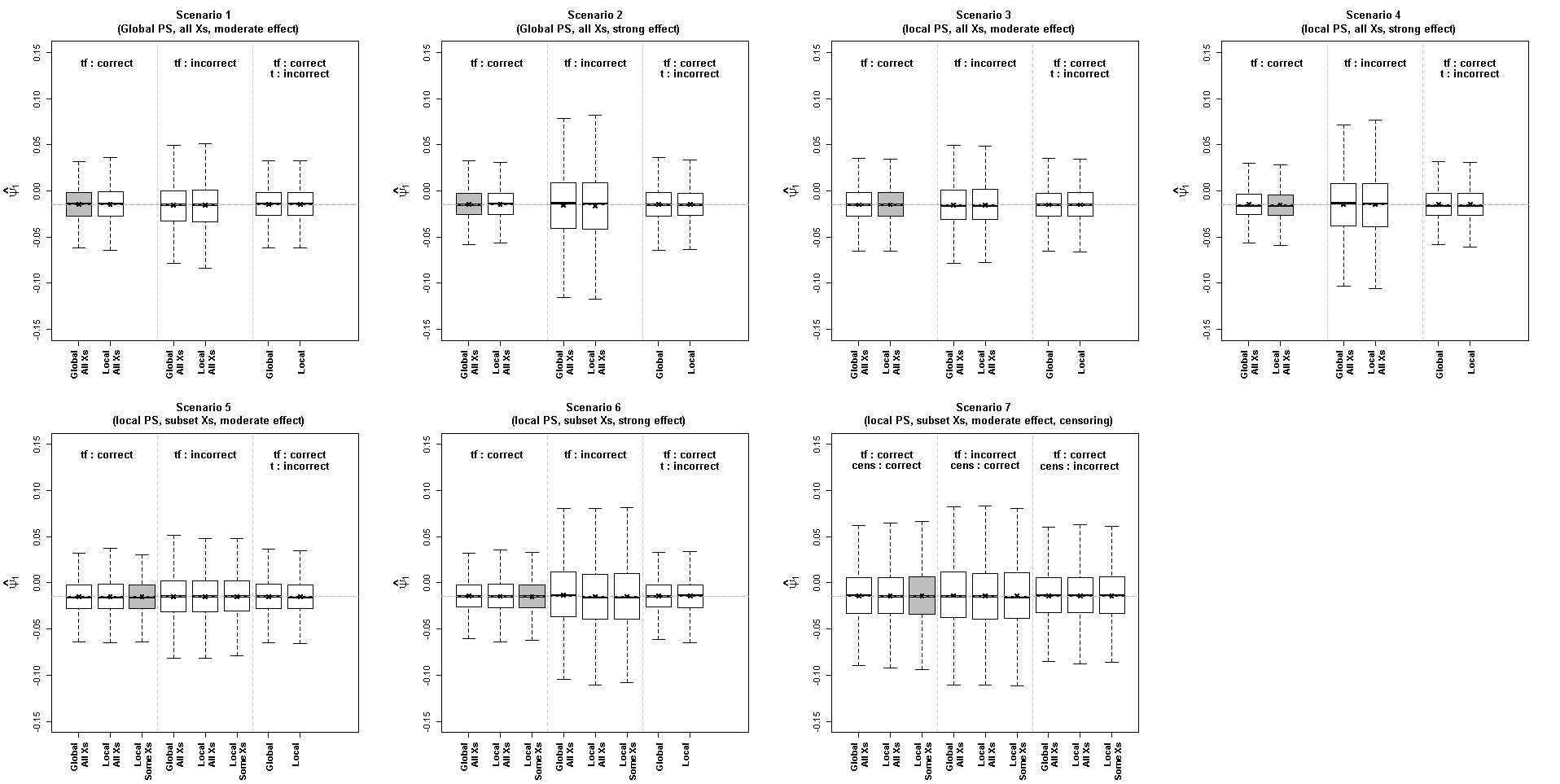}
        \caption{Simulation results: small sample size and small treatment effect settings - Performance of the methods over 1000 iterations in the estimation of $\psi_{1}$. For each scenario, the grey filling indicates the configuration under which the data have been simulated and the light-grey band represents the interval in which the absolute relative bias is less than or equal to 5\%. `PS' denotes propensity score, `tf' denotes treatment-free model, and 'cens' denotes censoring model. }
    \end{center}
\end{figure}
\end{landscape}

\scriptsize

\begin{table}\label{tab:SmallNLowTmt}
{
\caption{Simulation results: Simulations with small sample size and large treatment effect: Performance of the methods over 1000 iterations: mean estimate (95\%CI), relative bias (\%, denoted RB) and root mean squared error (RMSE) of $\hat\psi_0$, and difference in value function (dVF) between true and estimated ITR with its standard error (SE). Sc.~denotes scenario. }
\begin{tabular}{lllllll}
\hline
{Method} & Sc. & {Mean [CI]} & {RB}  & {RMSE} &  & dVF (SE) \\
\hline
&  & \multicolumn{5}{c}{\scriptsize{\textbf{Treatment-free model correctly specified, propensity score}}}\\
&  & \multicolumn{5}{c}{\scriptsize{\textbf{relies on all or a subset of all possible confounders}}} \\
\hline
Global All $X$s & 1 & 4.000 [3.659;4.342] & 0.012 & 0.030   & & $-3.20\times10^{-4}(4.40\times10^{-4})$\\
Local All $X$s & 1 & 4.000 [3.511;4.490] & 0.010 & 0.031   & & $-3.30\times10^{-4}(4.50\times10^{-4})$ \\
\hline
Global All $X$s & 2 & 4.001 [3.684;4.318] & 0.024 & 0.028   & & $-5.10\times10^{-4}(7.50\times10^{-4})$ \\
Local All $X$s & 2 & 4.000 [3.304;4.695] & -0.010 & 0.030   & & $-5.20\times10^{-4}(7.70\times10^{-4})$ \\
\hline
Global All $X$s & 3 & 3.997 [3.655;4.338] & -0.085 & 0.028  & & $-3.20\times10^{-4}(4.80\times10^{-4})$ \rule{0pt}{8pt}\\
Local All $X$s & 3 & 3.997 [3.506;4.488] & -0.077 & 0.030  & & $-3.20\times10^{-4}(4.90\times10^{-4})$ \rule{0pt}{8pt}\\
\hline
Global All $X$s & 4 & 4.002 [3.679;4.324] & 0.043 & 0.027 & &$-5.10\times10^{-4}(7.10\times10^{-4})$ \rule{0pt}{8pt}\\
Local All $X$s & 4 & 4.002 [3.336;4.668] & 0.060 & 0.029  & & $-5.10\times10^{-4}(7.30\times10^{-4})$ \rule{0pt}{8pt}\\
\hline
Global All $X$s & 5 & 4.003 [3.661;4.345] & 0.066 & 0.029  & & $-3.10\times10^{-4}(4.60\times10^{-4})$ \rule{0pt}{8pt}\\
Local All $X$s & 5 & 4.002 [3.512;4.491] & 0.039 & 0.030& & $-3.20\times10^{-4}(4.90\times10^{-4})$ \rule{0pt}{8pt}\\
Local Some $X$s & 5 & 4.003 [3.516;4.490] & 0.074 & 0.030  & & $-3.20\times10^{-4}(4.80\times10^{-4})$ \rule{0pt}{8pt}\\
\hline
Global All $X$s & 6 & 4.002 [3.663;4.342] & 0.056 & 0.028 & & $-5.20\times10^{-4}(7.40\times10^{-4})$ \rule{0pt}{8pt}\\
Local All $X$s & 6 & 4.002 [3.291;4.713] & 0.055 & 0.034  & & $-5.60\times10^{-4}(8.20\times10^{-4})$ \rule{0pt}{8pt}\\
Local Some $X$s & 6 & 4.002 [3.294;4.710] & 0.051 & 0.033  & & $-5.60\times10^{-4}(8.30\times10^{-4})$ \rule{0pt}{8pt}\\
\hline
Glob All $X$s and simcens & 7 & 4.006 [3.502;4.510] & 0.154 & 0.072  & & $-7.10\times10^{-4}(9.70\times10^{-4})$ \rule{0pt}{8pt}\\
Loc All $X$s and simcens & 7 & 4.006 [3.515;4.497] & 0.151 & 0.079  & & $-7.70\times10^{-4}(1.11\times10^{-3})$ \rule{0pt}{8pt}\\
Loc Some $X$s and simcens & 7 & 4.007 [3.519;4.495] & 0.176 & 0.078  & & $-7.60\times10^{-4}(1.10\times10^{-3})$ \rule{0pt}{8pt}\\

\hline
&  & \multicolumn{5}{c}{\scriptsize{\textbf{Treatment-free model misspecified, propensity score relies on  }}} \\
&  & \multicolumn{5}{c}{\scriptsize{\textbf{all or a subset of all possible confounders}}} \\
\hline
Global All $X$s & 1 & 4.003 [3.564;4.443] & 0.082 & 0.054   & & $-4.10\times10^{-4}(5.70\times10^{-4})$ \rule{0pt}{8pt}\\
Local All $X$s & 1 & 4.004 [3.374;4.633] & 0.099 & 0.055 & & $-4.20\times10^{-4}(5.90\times10^{-4})$ \rule{0pt}{8pt}\\
\hline
Global All $X$s & 2 & 4.007 [3.588;4.426] & 0.170 & 0.110  & & $-7.60\times10^{-4}(1.05\times10^{-3})$ \rule{0pt}{8pt}\\
Local All $X$s & 2 & 4.009 [3.090;4.929] & 0.237 & 0.112  & &  $-7.70\times10^{-4}(1.05\times10^{-3})$\rule{0pt}{8pt}\\
\hline
Global All $X$s & 3 & 4.006 [3.566;4.446] & 0.145 & 0.053  & & $-4.20\times10^{-4}(6.10\times10^{-4})$ \rule{0pt}{8pt}\\
Local All $X$s & 3 & 4.006 [3.375;4.638] & 0.158 & 0.054  & & $-4.20\times10^{-4}(6.30\times10^{-4})$ \rule{0pt}{8pt}\\
\hline
Global All $X$s & 4 & 4.008 [3.584;4.432] & 0.190 & 0.093  & & $-6.90\times10^{-4}(9.50\times10^{-4})$ \rule{0pt}{8pt}\\
Local All $X$s & 4 & 4.012 [3.137;4.888] & 0.308 & 0.098 & & $-7.20\times10^{-4}(9.90\times10^{-4})$ \rule{0pt}{8pt}\\
\hline
Global All $X$s & 5 & 4.000 [3.560;4.441] &  0.010 & 0.049  & & $-3.90\times10^{-4}(5.80\times10^{-4})$ \rule{0pt}{8pt}\\
Local All $X$s & 5 & 3.999 [3.371;4.628] & -0.018 & 0.049  & & $-4.10\times10^{-4}(6.10\times10^{-4})$ \rule{0pt}{8pt}\\
Local Some $X$s & 5 & 4.001 [3.375;4.626] & 0.013 & 0.049  & & $-4.00\times10^{-4}(6.10\times10^{-4})$ \rule{0pt}{8pt}\\
\hline
Global All $X$s & 6 & 3.994 [3.557;4.431] & -0.158 & 0.109  & & $-8.70\times10^{-4}(1.23\times10^{-3})$ \rule{0pt}{8pt}\\
Local All $X$s & 6 & 4.008 [3.084;4.931] & 0.192 & 0.116  & & $-9.10\times10^{-4}(1.32\times10^{-3})$ \rule{0pt}{8pt}\\
Local Some $X$s & 6 & 4.005 [3.086;4.924] & 0.136 & 0.115  & & $-9.00\times10^{-4}(1.33\times10^{-3})$ \rule{0pt}{8pt}\\
\hline
Glob All $X$s and simcens & 7 & 3.999 [3.364;4.634] & -0.016 & 0.116   & & $-9.50\times10^{-4}(1.39\times10^{-3})$ \rule{0pt}{8pt}\\
Loc All $X$s and simcens & 7 & 4.002 [3.371;4.632] & 0.040 & 0.127  & & $-1.02\times10^{-3}(1.59\times10^{-3})$ \rule{0pt}{8pt}\\
Loc Some $X$s and simcens & 7 & 4.002 [3.375;4.629] & 0.046 & 0.125  & & $-1.01\times10^{-3}(1.59\times10^{-3})$ \rule{0pt}{8pt}\\
\hline
&  & \multicolumn{5}{c}{\scriptsize{\textbf{Treatment-free model correctly specified, propensity score or }}} \\
&  & \multicolumn{5}{c}{\scriptsize{\textbf{censoring models misspecified (assumed independent of $X$s)}}} \\
\hline
Global & 1 & 4.001 [3.658;4.343] & 0.013 & 0.030 & & $-3.20\times10^{-4}(4.40\times10^{-4})$ \rule{0pt}{8pt}\\
Local & 1 & 4.000 [3.513;4.487] & 0.002 & 0.031   & & $-3.20\times10^{-4}(4.40\times10^{-4})$ \rule{0pt}{8pt}\\
\hline
Global & 2 & 4.001 [3.657;4.345] & 0.036 & 0.031   & & $-5.20\times10^{-4}(7.60\times10^{-4})$ \rule{0pt}{8pt}\\
Local & 2 & 4.001 [3.273;4.729] & 0.033 & 0.030  & & $-5.20\times10^{-4}(7.60\times10^{-4})$ \rule{0pt}{8pt}\\
\hline
Global & 3 & 3.997 [3.653;4.340] & -0.082 & 0.028   & & $-3.20\times10^{-4}(4.80\times10^{-4})$ \rule{0pt}{8pt}\\
Local & 3 & 3.997 [3.508;4.485] & -0.079 & 0.029   & & $-3.20\times10^{-4}(4.70\times10^{-4})$ \rule{0pt}{8pt}\\
\hline
Global & 4 & 4.000 [3.656;4.345] & 0.012 & 0.028   & & $-5.10\times10^{-4}(7.20\times10^{-4})$ \rule{0pt}{8pt}\\
Local & 4 & 4.001 [3.310;4.693] & 0.032 & 0.029   & & $-5.20\times10^{-4}(7.40\times10^{-4})$ \rule{0pt}{8pt}\\
\hline
Global & 5 & 4.003 [3.661;4.345] & 0.071 & 0.029   & & $-3.10\times10^{-4}(4.60\times10^{-4})$ \rule{0pt}{8pt}\\
Local & 5 & 4.003 [3.518;4.489] & 0.081 & 0.030   & & $-3.20\times10^{-4}(4.70\times10^{-4})$ \rule{0pt}{8pt}\\
\hline
Global & 6 & 4.003 [3.661;4.346] & 0.086 & 0.028  & & $-5.20\times10^{-4}(7.40\times10^{-4})$ \rule{0pt}{8pt}\\
Local & 6 & 4.003 [3.279;4.727] & 0.078 & 0.033   & &  $-5.50\times10^{-4}(8.20\times10^{-4})$\rule{0pt}{8pt}\\
\hline
Glob All $X$s and randcens & 7 & 4.006 [3.503;4.509] & 0.150 & 0.072   & & $-7.00\times10^{-4}(9.70\times10^{-4})$ \rule{0pt}{8pt}\\
Loc All $X$s and randcens & 7 & 4.005 [3.514;4.496] & 0.127 & 0.076  & & $-7.40\times10^{-4}(1.05\times10^{-3})$ \rule{0pt}{8pt}\\
Loc Some $X$s and randcens & 7 & 4.006 [3.517;4.495] & 0.155 & 0.075  & & $-7.30\times10^{-4}(1.05\times10^{-3})$ \rule{0pt}{8pt}\\

\hline
\end{tabular}
}

\parbox{18.5cm}{ ``Global All $X$s'' = Global propensity score with all possible confounders}
\parbox{18.5cm}{ ``Local All $X$s'' = Local propensity score with all possible confounders}
\parbox{18.5cm}{ ``Local Some $X$s'' = Local propensity score with known local confounders}
\parbox{18.5cm}{ ``Global'' = Global propensity score with treatment assumed to be randomly allocated}
\parbox{18.5cm}{ ``Local'' = Local propensity score with treatment assumed to be randomly allocated}
\end{table}

\small 

\begin{landscape}
\begin{figure}
   \begin{center}
        \includegraphics[scale=0.28]{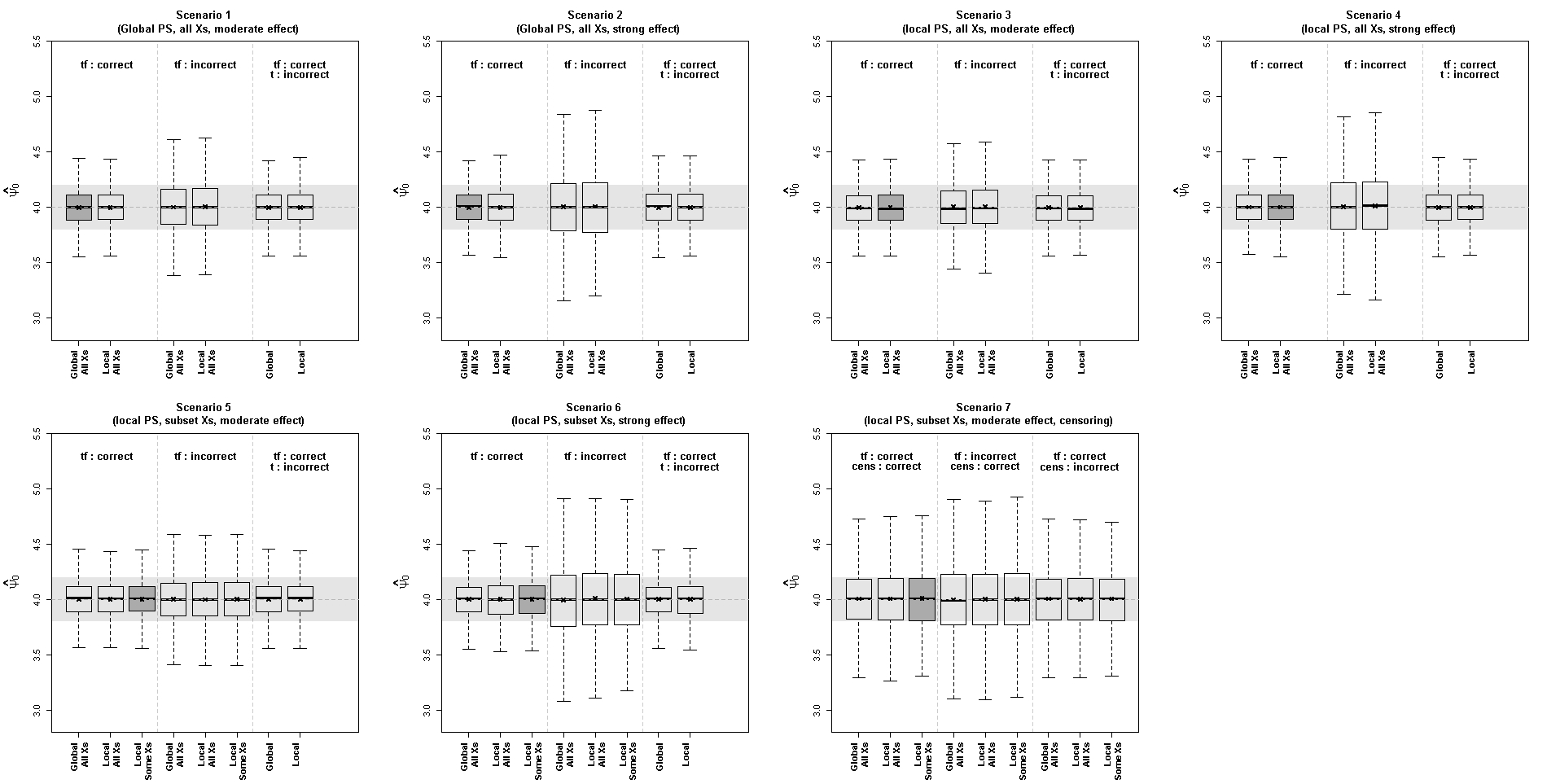}
        \caption{Simulation results: small sample size and large treatment effect settings - Performance of the methods over 1000 iterations in the estimation of $\psi_{0}$. For each scenario, the grey filling indicates the configuration under which the data have been simulated and the light-grey band represents the interval in which the absolute relative bias is less than or equal to 5\%. `PS' denotes propensity score, `tf' denotes treatment-free model, and 'cens' denotes censoring model.}
    \end{center}
\end{figure}
\end{landscape}

\scriptsize
\begin{table}\label{tab:SmallNLowTmt}
{
\caption{Simulation results: Simulations with small sample size and large treatment effect: Performance of the methods over 1000 iterations: mean estimate (95\%CI), relative bias (\%, denoted RB) and root mean squared error (RMSE) of $\hat\psi_1$, and difference in value function (dVF) between true and estimated ITR with its standard error (SE). Sc.~denotes scenario. }
\begin{tabular}{llllllll}
\hline
{Method} & Sc. & & & & {Mean [CI]} & {RB}  & {RMSE}  \\
\hline
&  & \multicolumn{6}{c}{\scriptsize{\textbf{Treatment-free model correctly specified, propensity score}}}\\
&  & \multicolumn{6}{c}{\scriptsize{\textbf{relies on all or a subset of all possible confounders}}} \\
\hline
Global All $X$s & 1 & & & & -0.550 [-0.587;-0.513] & 0.014 & $3.60\times10^{-4}$  \rule{0pt}{8pt}\\
Local All $X$s & 1 & & & &  -0.550 [-0.603;-0.497] & 0.011 & $3.70\times10^{-4}$  \rule{0pt}{8pt}\\

\hline

Global All $X$s & 2 & & & &  -0.550 [-0.585;-0.515] & -0.019 & $3.10\times10^{-4}$ \rule{0pt}{8pt}\\
Local All $X$s & 2 & & & &  -0.550 [-0.627;-0.473] & -0.044 & $3.20\times10^{-4}$  \rule{0pt}{8pt}\\

\hline

Global All $X$s & 3 & & & &  -0.550 [-0.587;-0.513] & -0.049 & $3.30\times10^{-4}$  \rule{0pt}{8pt}\\
Local All $X$s & 3 & & & &  -0.550 [-0.603;-0.496] & -0.042 & $3.50\times10^{-4}$  \rule{0pt}{8pt}\\

\hline

Global All $X$s & 4 & & & &  -0.550 [-0.586;-0.515] & 0.034 & $2.90\times10^{-4}$  \rule{0pt}{8pt}\\
Local All $X$s & 4 & & & &  -0.550 [-0.624;-0.477] & 0.064 & $3.20\times10^{-4}$  \rule{0pt}{8pt}\\

\hline

Global All $X$s & 5 & & & &  -0.550 [-0.587;-0.513] & 0.040 & $3.40\times10^{-4}$  \rule{0pt}{8pt}\\
Local All $X$s & 5 & & & &  -0.550 [-0.603;-0.497] & 0.013 & $3.40\times10^{-4}$  \rule{0pt}{8pt}\\
Local Some $X$s & 5 & & & &  -0.550 [-0.603;-0.497] & 0.045 & $3.40\times10^{-4}$  \rule{0pt}{8pt}\\

\hline

Global All $X$s & 6 & & & &  -0.550 [-0.587;-0.513] & 0.029 & $3.00\times10^{-4}$  \rule{0pt}{8pt}\\
Local All $X$s & 6 & & & &  -0.550 [-0.628;-0.473] & 0.036 & $3.80\times10^{-4}$  \rule{0pt}{8pt}\\
Local Some $X$s & 6 & & & &  -0.550 [-0.627;-0.473] & 0.029 & $3.60\times10^{-4}$  \rule{0pt}{8pt}\\

\hline

Glob All $X$s and simcens & 7 & & & &  -0.551 [-0.604;-0.497] & 0.103 & $8.10\times10^{-4}$ \rule{0pt}{8pt}\\
Loc All $X$s and simcens & 7 & & & &  -0.551 [-0.604;-0.497] & 0.109 & $9.00\times10^{-4}$  \rule{0pt}{8pt}\\
Loc Some $X$s and simcens & 7 & & & &  -0.551 [-0.604;-0.498] & 0.122 & $8.80\times10^{-4}$  \rule{0pt}{8pt}\\

\hline
&  & \multicolumn{6}{c}{\scriptsize{\textbf{Treatment-free model misspecified, propensity score relies on  }}} \\
&  & \multicolumn{6}{c}{\scriptsize{\textbf{all or a subset of all possible confounders}}} \\
\hline
Global All $X$s & 1 & & & &  -0.550 [-0.599;-0.502] & 0.070 & $6.60\times10^{-4}$  \rule{0pt}{8pt}\\
Local All $X$s & 1 & & & &  -0.550 [-0.620;-0.481] & 0.082 & $6.70\times10^{-4}$  \rule{0pt}{8pt}\\

\hline

Global All $X$s & 2 & & & &  -0.551 [-0.598;-0.503] & 0.102 & $1.39\times10^{-3}$ \rule{0pt}{8pt}\\
Local All $X$s & 2 & & & &  -0.551 [-0.654;-0.447] & 0.154 & $1.41\times10^{-3}$  \rule{0pt}{8pt}\\

\hline

Global All $X$s & 3 & & & &  -0.551 [-0.599;-0.502] & 0.135 & $6.40\times10^{-4}$  \rule{0pt}{8pt}\\
Local All $X$s & 3 & & & &  -0.551 [-0.620;-0.481] & 0.149 & $6.50\times10^{-4}$  \rule{0pt}{8pt}\\

\hline

Global All $X$s & 4 & & & &  -0.551 [-0.598;-0.503] & 0.158 & $1.19\times10^{-3}$  \rule{0pt}{8pt}\\
Local All $X$s & 4 & & & &  -0.551 [-0.650;-0.453] & 0.266 & $1.24\times10^{-3}$  \rule{0pt}{8pt}\\

\hline

Global All $X$s & 5 & & & &  -0.550 [-0.598;-0.502] & -0.004 & $5.80\times10^{-4}$  \rule{0pt}{8pt}\\
Local All $X$s & 5 & & & &  -0.550 [-0.619;-0.481] & -0.033 & $5.80\times10^{-4}$ \rule{0pt}{8pt}\\
Local Some $X$s & 5 & & & &  -0.550 [-0.619;-0.481] & -0.005 & $5.90\times10^{-4}$  \rule{0pt}{8pt}\\

\hline

Global All $X$s & 6 & & & &  -0.549 [-0.597;-0.501] & -0.121 & $1.30\times10^{-3}$  \rule{0pt}{8pt}\\
Local All $X$s & 6 & & & &  -0.551 [-0.653;-0.449] & 0.140 & $1.39\times10^{-3}$  \rule{0pt}{8pt}\\
Local Some $X$s & 6 & & & &  -0.551 [-0.652;-0.449] & 0.093 & $1.37\times10^{-3}$ \rule{0pt}{8pt}\\

\hline

Glob All $X$s and simcens & 7 & & & &  -0.550 [-0.618;-0.482] & -0.030 & $1.33\times10^{-3}$  \rule{0pt}{8pt}\\
Loc All $X$s and simcens & 7 & & & &  -0.550 [-0.619;-0.481] & 0.018 &  $1.46\times10^{-3}$  \rule{0pt}{8pt}\\
Loc Some $X$s and simcens & 7 & & & &  -0.550 [-0.619;-0.481] &  0.017  & $1.43\times10^{-3}$ \rule{0pt}{8pt}\\

\hline
&  & \multicolumn{6}{c}{\scriptsize{\textbf{Treatment-free model correctly specified, propensity score or }}} \\
&  & \multicolumn{6}{c}{\scriptsize{\textbf{censoring models misspecified (assumed independent of $X$s)}}} \\
\hline
Global & 1 & & & &  -0.550 [-0.587;-0.513] & 0.015 & $3.60\times10^{-4}$\rule{0pt}{8pt}\\
Local & 1 & & & &  -0.550 [-0.603;-0.497] & 0.006 & $3.60\times10^{-4}$ \rule{0pt}{8pt}\\

\hline

Global & 2 & & & &  -0.550 [-0.587;-0.512] & -0.007 & $3.40\times10^{-4}$ \rule{0pt}{8pt}\\
Local & 2 & & & &  -0.550 [-0.630;-0.470] & -0.010 & $3.30\times10^{-4}$ \rule{0pt}{8pt}\\

\hline

Global & 3 & & & &  -0.550 [-0.587;-0.512] & -0.047 & $3.30\times10^{-4}$ \rule{0pt}{8pt}\\
Local & 3 & & & &  -0.550 [-0.603;-0.497] & -0.044 & $3.40\times10^{-4}$ \rule{0pt}{8pt}\\

\hline

Global & 4 & & & &  -0.550 [-0.588;-0.513] & 0.009 & $3.20\times10^{-4}$ \rule{0pt}{8pt}\\
Local & 4 & & & &  -0.550 [-0.626;-0.474] & 0.026 & $3.30\times10^{-4}$ \rule{0pt}{8pt}\\

\hline

Global & 5 & & & &  -0.550 [-0.587;-0.513] & 0.044 & $3.40\times10^{-4}$ \rule{0pt}{8pt}\\
Local & 5 & & & &  -0.550 [-0.603;-0.498] & 0.050 & $3.40\times10^{-4}$ \rule{0pt}{8pt}\\

\hline

Global & 6 & & & &  -0.550 [-0.587;-0.513] & 0.050 & $3.00\times10^{-4}$ \rule{0pt}{8pt}\\
Local & 6 & & & &  -0.550 [-0.629;-0.472] & 0.047 & $3.60\times10^{-4}$ \rule{0pt}{8pt}\\

\hline

Glob All $X$s and randcens & 7 & & & &  -0.551 [-0.604;-0.497] &  0.101 & $8.10\times10^{-4}$ \rule{0pt}{8pt}\\
Loc All $X$s and randcens & 7 & & & &  -0.551 [-0.604;-0.497] & 0.097 & $8.60\times10^{-4}$ \rule{0pt}{8pt}\\
Loc Some $X$s and randcens & 7 & & & &  -0.551 [-0.604;-0.498] & 0.113 & $8.50\times10^{-4}$ \rule{0pt}{8pt}\\

\hline
\end{tabular}
}

\parbox{18.5cm}{ ``Global All $X$s'' = Global propensity score with all possible confounders}
\parbox{18.5cm}{ ``Local All $X$s'' = Local propensity score with all possible confounders}
\parbox{18.5cm}{ ``Local Some $X$s'' = Local propensity score with known local confounders}
\parbox{18.5cm}{ ``Global'' = Global propensity score with treatment assumed to be randomly allocated}
\parbox{18.5cm}{ ``Local'' = Local propensity score with treatment assumed to be randomly allocated}
\end{table}

\begin{landscape}
\begin{figure}
   \begin{center}
        \includegraphics[scale=0.28]{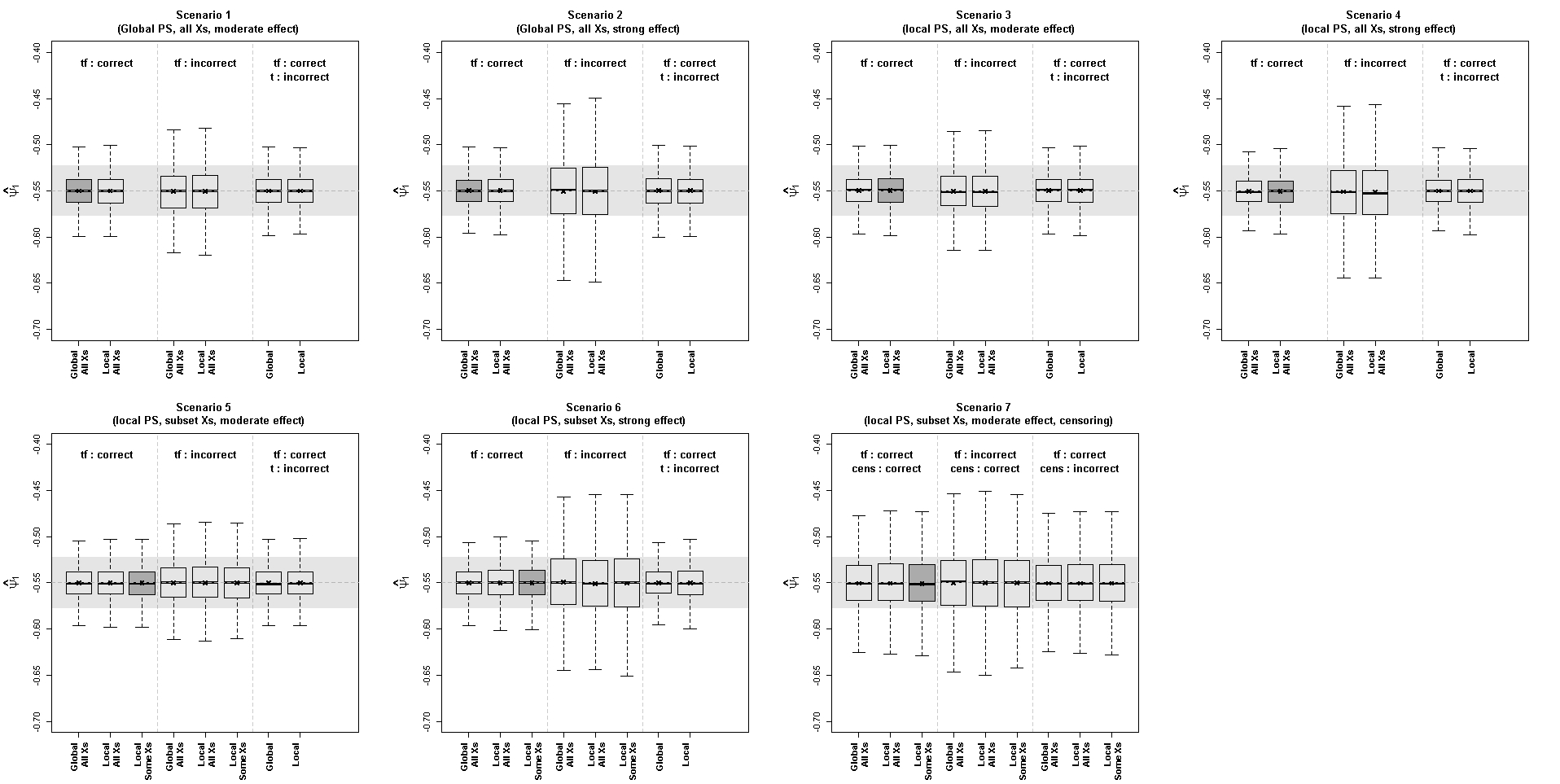}
        \caption{Simulation results: small sample size and large treatment effect settings - Performance of the methods over 1000 iterations in the estimation of $\psi_{1}$. For each scenario, the grey filling indicates the configuration under which the data have been simulated and the light-grey band represents the interval in which the absolute relative bias is less than or equal to 5\%. `PS' denotes propensity score, `tf' denotes treatment-free model, and 'cens' denotes censoring model.}
    \end{center}
\end{figure}
\end{landscape}



\begin{table}\label{tab:SmallNLowTmt}
{
\caption{Simulation results: Simulations with a large sample size and small treatment effect: Performance of the methods over 1000 iterations: mean estimate (95\%CI), relative bias (\%, denoted RB) and root mean squared error (RMSE) of $\hat\psi_0$, and difference in value function (dVF) between true and estimated ITR with its standard error (SE). Sc.~denotes scenario. }
\begin{tabular}{lllllll}
\hline
{Method} & Sc. & {Mean [CI]} & {RB}  & {RMSE} &  & dVF (SE) \\
\hline
&  & \multicolumn{5}{c}{\scriptsize{\textbf{Treatment-free model correctly specified, propensity score}}}\\
&  & \multicolumn{5}{c}{\scriptsize{\textbf{relies on all or a subset of all possible confounders}}} \\
\hline
Global All $X$s & 1 & 0.151 [0.117;0.185] & 0.615 & $2.80\times10^{-4}$   & & $-1.50\times10^{-4}(2.20\times10^{-4})$\rule{0pt}{8pt}\\
Local All $X$s & 1 & 0.151 [0.102;0.199] & 0.617 & $2.80\times10^{-4}$   & & $-1.50\times10^{-4}(2.20\times10^{-4})$\rule{0pt}{8pt}\\

\hline

Global All $X$s & 2 & 0.150 [0.116;0.184] & 0.204 & $2.70\times10^{-4}$  & & $-1.70\times10^{-4}(2.40\times10^{-4})$\rule{0pt}{8pt}\\
Local All $X$s & 2 & 0.150 [0.102;0.199] & 0.212 & $2.70\times10^{-4}$  & & $-1.70\times10^{-4}(2.40\times10^{-4})$\rule{0pt}{8pt}\\

\hline

Global All $X$s & 3 & 0.150 [0.116;0.184] & -0.105 & $2.90\times10^{-4}$   & & $-1.50\times10^{-4}(2.20\times10^{-4})$\rule{0pt}{8pt}\\
Local All $X$s & 3 & 0.150 [0.101;0.198] & -0.103 & $2.90\times10^{-4}$   & & $-1.50\times10^{-4}(2.20\times10^{-4})$\rule{0pt}{8pt}\\

\hline

Global All $X$s & 4 & 0.150 [0.116;0.184] & 0.107 & $3.00\times10^{-4}$  & & $-1.60\times10^{-4}(2.40\times10^{-4})$\rule{0pt}{8pt}\\
Local All $X$s & 4 & 0.150 [0.102;0.199] & 0.105 & $3.00\times10^{-4}$ & & $-1.60\times10^{-4}(2.40\times10^{-4})$\rule{0pt}{8pt}\\

\hline

Global All $X$s & 5 & 0.150 [0.116;0.184] & -0.176 & $2.80\times10^{-4}$  & & $-1.70\times10^{-4}(2.10\times10^{-4})$\rule{0pt}{8pt}\\
Local All $X$s & 5 & 0.150 [0.101;0.198] & -0.173 & $2.80\times10^{-4}$ & & $-1.70\times10^{-4}(2.10\times10^{-4})$\rule{0pt}{8pt}\\
Local Some $X$s & 5 & 0.150 [0.101;0.198] & -0.177 & $2.80\times10^{-4}$ & & $-1.70\times10^{-4}(2.10\times10^{-4})$\rule{0pt}{8pt}\\

\hline

Global All $X$s & 6 & 0.150 [0.116;0.184] & 0.111 & $3.00\times10^{-4}$ & & $-1.60\times10^{-4}(2.20\times10^{-4})$\rule{0pt}{8pt}\\
Local All $X$s & 6 & 0.150 [0.102;0.199] & 0.105 & $3.00\times10^{-4}$  & & $-1.60\times10^{-4}(2.20\times10^{-4})$\rule{0pt}{8pt}\\
Local Some $X$s & 6 & 0.150 [0.102;0.199] & 0.111 & $3.00\times10^{-4}$  & & $-1.60\times10^{-4}(2.20\times10^{-4})$\rule{0pt}{8pt}\\

\hline

Global& 7 & 0.150 [0.100;0.200] & 0.140 & $6.20\times10^{-4}$ & & $-2.80\times10^{-4}(3.90\times10^{-4})$\rule{0pt}{8pt}\\
Local All $X$s  & 7 & 0.150 [0.102;0.198] & 0.043 & $6.50\times10^{-4}$ & & $-3.10\times10^{-4}(4.20\times10^{-4})$\rule{0pt}{8pt}\\
Local Some $X$s  & 7 & 0.150 [0.102;0.198] & 0.037 & $6.50\times10^{-4}$  & & $-3.10\times10^{-4}(4.20\times10^{-4})$\rule{0pt}{8pt}\\

\hline
&  & \multicolumn{5}{c}{\scriptsize{\textbf{Treatment-free model misspecified, propensity score relies on  }}} \\
&  & \multicolumn{5}{c}{\scriptsize{\textbf{all or a subset of all possible confounders}}} \\
\hline
Global All $X$s & 1 & 0.151 [0.107;0.195] & 0.831 &  $5.20\times10^{-4}$  & & $-1.80\times10^{-4}(2.80\times10^{-4})$\rule{0pt}{8pt}\\
Local All $X$s & 1 & 0.151 [0.089;0.214] & 0.854 &  $5.20\times10^{-4}$  & & $-1.80\times10^{-4}(2.80\times10^{-4})$\rule{0pt}{8pt}\\

\hline

Global All $X$s & 2 & 0.151 [0.107;0.195] & 0.467 & $5.10\times10^{-4}$ & & $-2.10\times10^{-4}(3.60\times10^{-4})$\rule{0pt}{8pt}\\
Local All $X$s & 2 & 0.151 [0.088;0.213] & 0.469 & $5.10\times10^{-4}$   & & $-2.10\times10^{-4}(3.60\times10^{-4})$\rule{0pt}{8pt}\\

\hline

Global All $X$s & 3 & 0.149 [0.106;0.193] & -0.351 & $5.60\times10^{-4}$  & & $-1.90\times10^{-4}(2.80\times10^{-4})$\rule{0pt}{8pt}\\
Local All $X$s & 3 & 0.150 [0.087;0.212] &  -0.321 & $5.60\times10^{-4}$  & & $-1.90\times10^{-4}(2.80\times10^{-4})$\rule{0pt}{8pt}\\

\hline

Global All $X$s & 4 & 0.150 [0.106;0.194] & 0.214 & $5.00\times10^{-4}$  & & $-1.90\times10^{-4}(2.90\times10^{-4})$\rule{0pt}{8pt}\\
Local All $X$s & 4 & 0.150 [0.088;0.213] & 0.218 & $5.00\times10^{-4}$  & & $-1.90\times10^{-4}(2.90\times10^{-4})$\rule{0pt}{8pt}\\

\hline

Global All $X$s & 5 & 0.150 [0.106;0.194] & -0.249 & $5.00\times10^{-4}$  & & $-1.90\times10^{-4}(2.80\times10^{-4})$\rule{0pt}{8pt}\\
Local All $X$s & 5 & 0.150 [0.087;0.212] & -0.259 & $5.10\times10^{-4}$  & & $-1.90\times10^{-4}(2.80\times10^{-4})$\rule{0pt}{8pt}\\
Local Some $X$s & 5 & 0.147 [0.085;0.210] & -1.669 & $5.10\times10^{-4}$  & & $-2.00\times10^{-4}(2.90\times10^{-4})$\rule{0pt}{8pt}\\

\hline

Global All $X$s & 6 & 0.150 [0.106;0.193] & -0.309 & $5.70\times10^{-4}$  & & $-2.00\times10^{-4}(2.90\times10^{-4})$\rule{0pt}{8pt}\\
Local All $X$s & 6 & 0.150 [0.087;0.212] & -0.293 & $5.70\times10^{-4}$   & & $-2.00\times10^{-4}(2.90\times10^{-4})$\rule{0pt}{8pt}\\
Local Some $X$s & 6 & 0.147 [0.085;0.210] & -1.699 & $5.70\times10^{-4}$  & & $-2.10\times10^{-4}(3.00\times10^{-4})$\rule{0pt}{8pt}\\

\hline

Global& 7 & 0.150 [0.086;0.213] & -0.262 & $1.00\times10^{-3}$   & & $-3.20\times10^{-4}(4.70\times10^{-4})$\rule{0pt}{8pt}\\
Local All $X$s  & 7 & 0.150 [0.088;0.212] & -0.165 & $1.08\times10^{-3}$  & & $-3.50\times10^{-4}(4.90\times10^{-4})$\rule{0pt}{8pt}\\
Local Some $X$s  & 7 & 0.150 [0.088;0.212] & -0.162 & $1.08\times10^{-3}$   & & $-3.50\times10^{-4}(4.90\times10^{-4})$\rule{0pt}{8pt}\\
\hline
&  & \multicolumn{5}{c}{\scriptsize{\textbf{Treatment-free model correctly specified, propensity score or }}} \\
&  & \multicolumn{5}{c}{\scriptsize{\textbf{censoring models misspecified (assumed independent of $X$s)}}} \\
\hline
Global  & 1 & 0.151 [0.117;0.185] & 0.619 & $2.80\times10^{-4}$  & & $-1.50\times10^{-4}(2.20\times10^{-4})$\rule{0pt}{8pt}\\
Local  & 1 & 0.151 [0.102;0.200] & 0.618 & $2.80\times10^{-4}$  & & $-1.50\times10^{-4}(2.20\times10^{-4})$\rule{0pt}{8pt}\\

\hline

Global  & 2 & 0.150 [0.116;0.185] & 0.211 & $2.70\times10^{-4}$  & & $-1.70\times10^{-4}(2.40\times10^{-4})$\rule{0pt}{8pt}\\
Local  & 2 & 0.150 [0.102;0.199] & 0.212 & $2.70\times10^{-4}$ & & $-1.70\times10^{-4}(2.40\times10^{-4})$\rule{0pt}{8pt}\\

\hline

Global  & 3 & 0.150 [0.116;0.184] & -0.113 & $2.90\times10^{-4}$  & & $-1.50\times10^{-4}(2.20\times10^{-4})$\rule{0pt}{8pt}\\
Local  & 3 & 0.150 [0.101;0.198] & -0.114 & $2.90\times10^{-4}$  & & $-1.50\times10^{-4}(2.20\times10^{-4})$\rule{0pt}{8pt}\\

\hline

Global  & 4 & 0.150 [0.116;0.184] & 0.108 & $3.00\times10^{-4}$  & & $-1.60\times10^{-4}(2.30\times10^{-4})$\rule{0pt}{8pt}\\
Local  & 4 & 0.150 [0.102;0.199] & 0.106 & $3.00\times10^{-4}$  & & $-1.60\times10^{-4}(2.40\times10^{-4})$\rule{0pt}{8pt}\\

\hline

Global  & 5 & 0.150 [0.115;0.184] & -0.180 & $2.8\times10^{-4}$  & & $-1.70\times10^{-4}(2.10\times10^{-4})$\rule{0pt}{8pt}\\
Local  & 5 & 0.150 [0.101;0.198] & -0.179 & $2.8\times10^{-4}$  & & $-1.70\times10^{-4}(2.10\times10^{-4})$\rule{0pt}{8pt}\\

\hline

Global  & 6 & 0.150 [0.116;0.184] & 0.105 & $3.00\times10^{-4}$  & & $-1.60\times10^{-4}(2.20\times10^{-4})$\rule{0pt}{8pt}\\
Local  & 6 & 0.150 [0.102;0.199] & 0.105 & $3.00\times10^{-4}$  & & $-1.60\times10^{-4}(2.20\times10^{-4})$\rule{0pt}{8pt}\\

\hline

Global& 7 & 0.150 [0.100;0.200] & 0.144 & $6.20\times10^{-4}$   & & $-2.80\times10^{-4}(3.90\times10^{-4})$\rule{0pt}{8pt}\\
Local All $X$s  & 7 & 0.150 [0.102;0.199] & 0.111 & $6.30\times10^{-4}$  & & $-3.00\times10^{-4}(4.10\times10^{-4})$\rule{0pt}{8pt}\\
Local Some $X$s  & 7 & 0.150 [0.102;0.199] & 0.104 & $6.30\times10^{-4}$   & & $-3.00\times10^{-4}(4.10\times10^{-4})$\rule{0pt}{8pt}\\
\hline
\end{tabular}
}

\parbox{18.5cm}{ ``Global All $X$s'' = Global propensity score with all possible confounders}
\parbox{18.5cm}{ ``Local All $X$s'' = Local propensity score with all possible confounders}
\parbox{18.5cm}{ ``Local Some $X$s'' = Local propensity score with known local confounders}
\parbox{18.5cm}{ ``Global'' = Global propensity score with treatment assumed to be randomly allocated}
\parbox{18.5cm}{ ``Local'' = Local propensity score with treatment assumed to be randomly allocated}
\end{table}

\small

\begin{landscape}
\begin{figure}
   \begin{center}
        \includegraphics[scale=0.28]{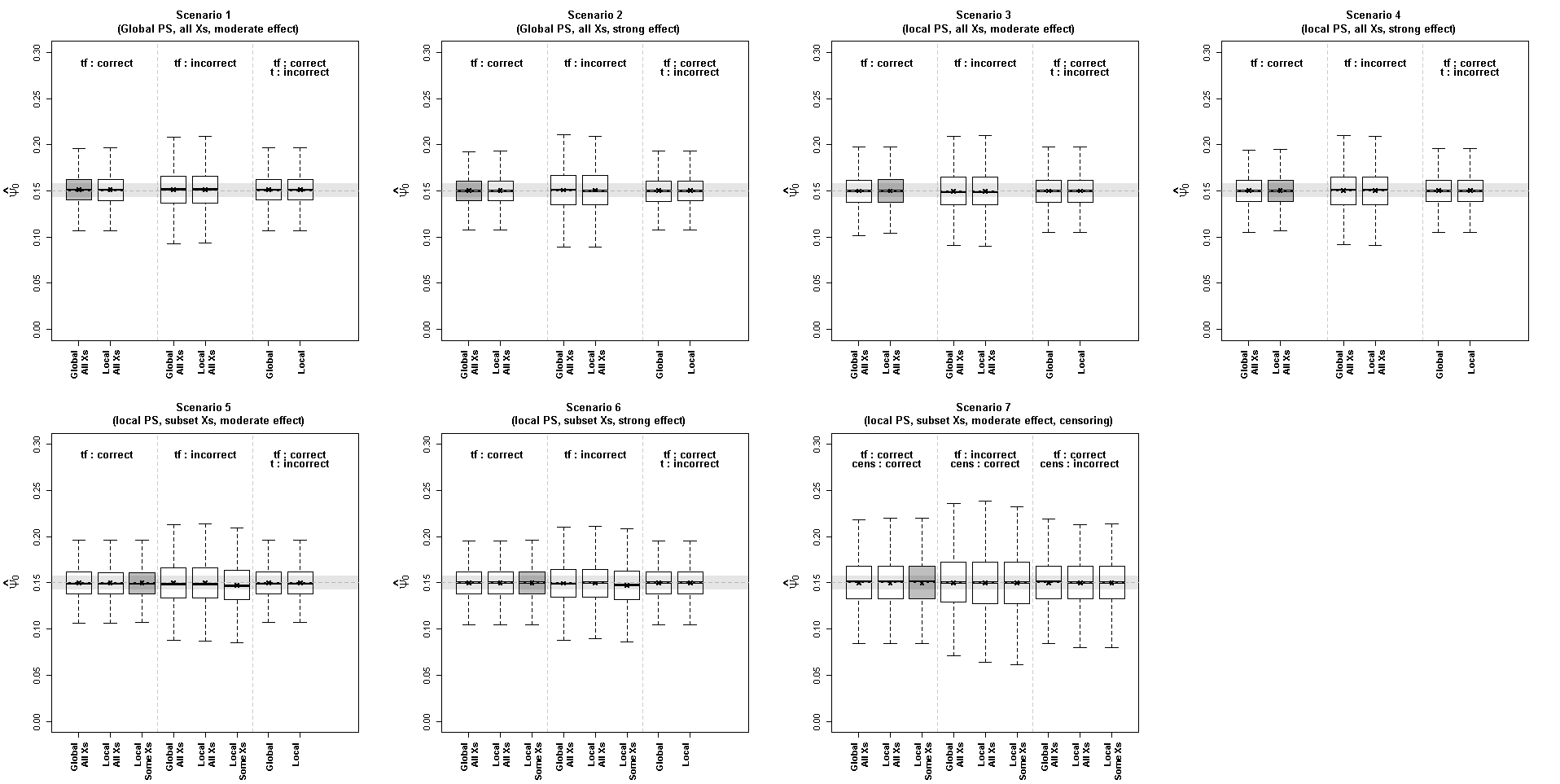}
        \caption{Simulation results: large sample size and small treatment effect settings - Performance of the methods over 1000 iterations in the estimation of $\psi_{0}$. For each scenario, the grey filling indicates the configuration under which the data have been simulated and the light-grey band represents the interval in which the absolute relative bias is less than or equal to 5\%. `PS' denotes propensity score, `tf' denotes treatment-free model, and 'cens' denotes censoring model.}
    \end{center}
\end{figure}
\end{landscape}

\scriptsize
\begin{table}\label{tab:SmallNLowTmt}
{
\caption{Simulation results: Simulations with a large sample size and small treatment effect: Performance of the methods over 1000 iterations: mean estimate (95\%CI), relative bias (\%, denoted RB) and root mean squared error (RMSE) of $\hat\psi_1$, and difference in value function (dVF) between true and estimated ITR with its standard error (SE). Sc.~denotes scenario. }
\begin{tabular}{llllllll}
\hline
{Method} & Sc. &&&& {Mean [CI]} & {RB}  & {RMSE}  \\
\hline
&  & \multicolumn{6}{c}{\scriptsize{\textbf{Treatment-free model correctly specified, propensity score}}}\\
&  & \multicolumn{6}{c}{\scriptsize{\textbf{relies on all or a subset of all possible confounders}}} \\
\hline
Global All $X$s & 1 &&&& -0.015 [-0.019;-0.011] & 0.568 & $3.20\times10^{-6}$ \rule{0pt}{8pt}\\
Local All $X$s & 1 &&&&-0.015 [-0.020;-0.010] & 0.570 & $3.20\times10^{-6}$ \rule{0pt}{8pt}\\

\hline

Global All $X$s & 2 &&&&  -0.015 [-0.019;-0.011] & 0.288 & $3.20\times10^{-6}$ \rule{0pt}{8pt}\\
Local All $X$s & 2 &&&&  -0.015 [-0.020;-0.010] & 0.297 & $3.20\times10^{-6}$ \rule{0pt}{8pt}\\

\hline

Global All $X$s & 3 &&&&  -0.015 [-0.019;-0.011] & -0.249 & $3.40\times10^{-6}$ \rule{0pt}{8pt}\\
Local All $X$s & 3 &&&&  -0.015 [-0.020;-0.010] & -0.247 & $3.40\times10^{-6}$ \rule{0pt}{8pt}\\

\hline

Global All $X$s & 4 &&&&  -0.015 [-0.019;-0.011] & -0.085 & $3.40\times10^{-6}$ \rule{0pt}{8pt}\\
Local All $X$s & 4 &&&&  -0.015 [-0.020;-0.010] & -0.088 & $3.40\times10^{-6}$\rule{0pt}{8pt}\\

\hline

Global All $X$s & 5 &&&&  -0.015 [-0.019;-0.011] & -0.280 & $3.30\times10^{-6}$ \rule{0pt}{8pt}\\
Local All $X$s & 5 &&&&  -0.015 [-0.020;-0.010] & -0.276  & $3.30\times10^{-6}$ \rule{0pt}{8pt}\\
Local Some $X$s & 5 &&&&  -0.015 [-0.020;-0.010] & -0.281 & $3.30\times10^{-6}$ \rule{0pt}{8pt}\\

\hline

Global All $X$s & 6 &&&&  -0.015 [-0.019;-0.011] & 0.104 & $3.40\times10^{-6}$ \rule{0pt}{8pt}\\
Local All $X$s & 6 &&&&  -0.015 [-0.020;-0.010] & 0.098 & $3.40\times10^{-6}$ \rule{0pt}{8pt}\\
Local Some $X$s & 6 &&&&  -0.015 [-0.020;-0.010] & 0.103 & $3.40\times10^{-6}$ \rule{0pt}{8pt}\\

\hline

Global& 7 &&&&  -0.015 [-0.020;-0.010] & -0.075 & $7.00\times10^{-6}$ \rule{0pt}{8pt}\\
Local All $X$s  & 7 &&&&  -0.015 [-0.020;-0.010] & -0.089 & $7.30\times10^{-6}$ \rule{0pt}{8pt}\\
Local Some $X$s  & 7 &&&&  -0.015 [-0.020;-0.010] & -0.095 & $7.30\times10^{-6}$ \rule{0pt}{8pt}\\

\hline
&  & \multicolumn{6}{c}{\scriptsize{\textbf{Treatment-free model misspecified, propensity score relies on  }}} \\
&  & \multicolumn{6}{c}{\scriptsize{\textbf{all or a subset of all possible confounders}}} \\
\hline
Global All $X$s & 1 &&&& -0.015 [-0.020;-0.010] & 0.815 & $6.10\times10^{-6}$ \rule{0pt}{8pt}\\
Local All $X$s & 1 &&&&  -0.015 [-0.022;-0.008] & 0.837 & $6.10\times10^{-6}$ \rule{0pt}{8pt}\\

\hline

Global All $X$s & 2 &&&&  -0.015 [-0.020;-0.010] & 0.582 & $6.30\times10^{-6}$\rule{0pt}{8pt}\\
Local All $X$s & 2 &&&&  -0.015 [-0.022;-0.008] & 0.586 & $6.30\times10^{-6}$ \rule{0pt}{8pt}\\

\hline

Global All $X$s & 3 &&&&  -0.015 [-0.020;-0.010] & -0.512 & $6.60\times10^{-6}$ \rule{0pt}{8pt}\\
Local All $X$s & 3 &&&&  -0.015 [-0.022;-0.008] & -0.485 & $6.60\times10^{-6}$ \rule{0pt}{8pt}\\

\hline

Global All $X$s & 4 &&&&   -0.015 [-0.020;-0.010] & 0.034 & $5.90\times10^{-6}$ \rule{0pt}{8pt}\\
Local All $X$s & 4 &&&& -0.015 [-0.022;-0.008] & 0.038 & $5.90\times10^{-6}$\rule{0pt}{8pt}\\

\hline

Global All $X$s & 5 &&&&  -0.015 [-0.020;-0.010] & -0.356 & $6.00\times10^{-6}$ \rule{0pt}{8pt}\\
Local All $X$s & 5 &&&&  -0.015 [-0.022;-0.008] & -0.365 & $6.00\times10^{-6}$ \rule{0pt}{8pt}\\
Local Some $X$s & 5 &&&&  -0.015 [-0.022;-0.008] & -1.872 & $6.10\times10^{-6}$\rule{0pt}{8pt}\\

\hline

Global All $X$s & 6 &&&&  -0.015 [-0.020;-0.010] & -0.370 & $6.70\times10^{-6}$ \rule{0pt}{8pt}\\
Local All $X$s & 6 &&&&  -0.015 [-0.022;-0.008] & -0.354 & $6.70\times10^{-6}$ \rule{0pt}{8pt}\\
Local Some $X$s & 6 &&&&  -0.015 [-0.022;-0.008] & -1.857 & $6.70\times10^{-6}$ \rule{0pt}{8pt}\\

\hline

Global& 7 &&&&  -0.015 [-0.022;-0.008] & -0.493 & $1.16\times10^{-5}$ \rule{0pt}{8pt}\\
Local All $X$s  & 7 &&&&  -0.015 [-0.022;-0.008] & -0.330 & $1.25\times10^{-5}$ \rule{0pt}{8pt}\\
Local Some $X$s  & 7 &&&&  -0.015 [-0.022;-0.008] & -0.325 & $1.25\times10^{-5}$ \rule{0pt}{8pt}\\

\hline
&  & \multicolumn{6}{c}{\scriptsize{\textbf{Treatment-free model correctly specified, propensity score or }}} \\
&  & \multicolumn{6}{c}{\scriptsize{\textbf{censoring models misspecified (assumed independent of $X$s)}}} \\
\hline
Global  & 1&&&&  -0.015 [-0.019;-0.011] & 0.572 & $3.20\times10^{-6}$ \rule{0pt}{8pt}\\
Local  & 1 &&&&  -0.015 [-0.020;-0.010] & 0.571 & $3.10\times10^{-6}$ \rule{0pt}{8pt}\\

\hline

Global  & 2 &&&&  -0.015 [-0.019;-0.011] & 0.296 & $3.20\times10^{-6}$ \rule{0pt}{8pt}\\
Local  & 2 &&&& -0.015 [-0.020;-0.010] & 0.297 & $3.20\times10^{-6}$ \rule{0pt}{8pt}\\

\hline

Global  & 3 &&&&  -0.015 [-0.019;-0.011] & -0.257 & $3.40\times10^{-6}$ \rule{0pt}{8pt}\\
Local  & 3 &&&&  -0.015 [-0.020;-0.010] & -0.260 & $3.40\times10^{-6}$ \rule{0pt}{8pt}\\

\hline

Global  & 4 &&&&  -0.015 [-0.019;-0.011] & -0.084 & $3.40\times10^{-6}$ \rule{0pt}{8pt}\\
Local  & 4 &&&&  -0.015 [-0.020;-0.010] & -0.087 & $3.40\times10^{-6}$\rule{0pt}{8pt}\\

\hline

Global  & 5 &&&&  -0.015 [-0.019;-0.011] & -0.284 & $3.30\times10^{-6}$ \rule{0pt}{8pt}\\
Local  & 5 &&&&  -0.015 [-0.020;-0.010] & -0.285 & $3.30\times10^{-6}$ \rule{0pt}{8pt}\\

\hline

Global  & 6 &&&&  -0.015 [-0.019;-0.011] & 0.099 & $3.40\times10^{-6}$ \rule{0pt}{8pt}\\
Local  & 6 &&&&  -0.015 [-0.020;-0.010] & 0.097 & $3.40\times10^{-6}$ \rule{0pt}{8pt}\\

\hline

Global& 7 &&&&  -0.015 [-0.020;-0.010] & -0.068 & $6.90\times10^{-6}$ \rule{0pt}{8pt}\\
Local All $X$s  & 7 &&&&  -0.015 [-0.020;-0.010] & -0.043 & $7.10\times10^{-6}$ \rule{0pt}{8pt}\\
Local Some $X$s  & 7 &&&&  -0.015 [-0.020;-0.010] & -0.049 & $7.10\times10^{-6}$ \rule{0pt}{8pt}\\
\hline
\end{tabular}
}

\parbox{18.5cm}{ ``Global All $X$s'' = Global propensity score with all possible confounders}
\parbox{18.5cm}{ ``Local All $X$s'' = Local propensity score with all possible confounders}
\parbox{18.5cm}{ ``Local Some $X$s'' = Local propensity score with known local confounders}
\parbox{18.5cm}{ ``Global'' = Global propensity score with treatment assumed to be randomly allocated}
\parbox{18.5cm}{ ``Local'' = Local propensity score with treatment assumed to be randomly allocated}
\end{table}

\begin{landscape}
\begin{figure}
   \begin{center}
        \includegraphics[scale=0.28]{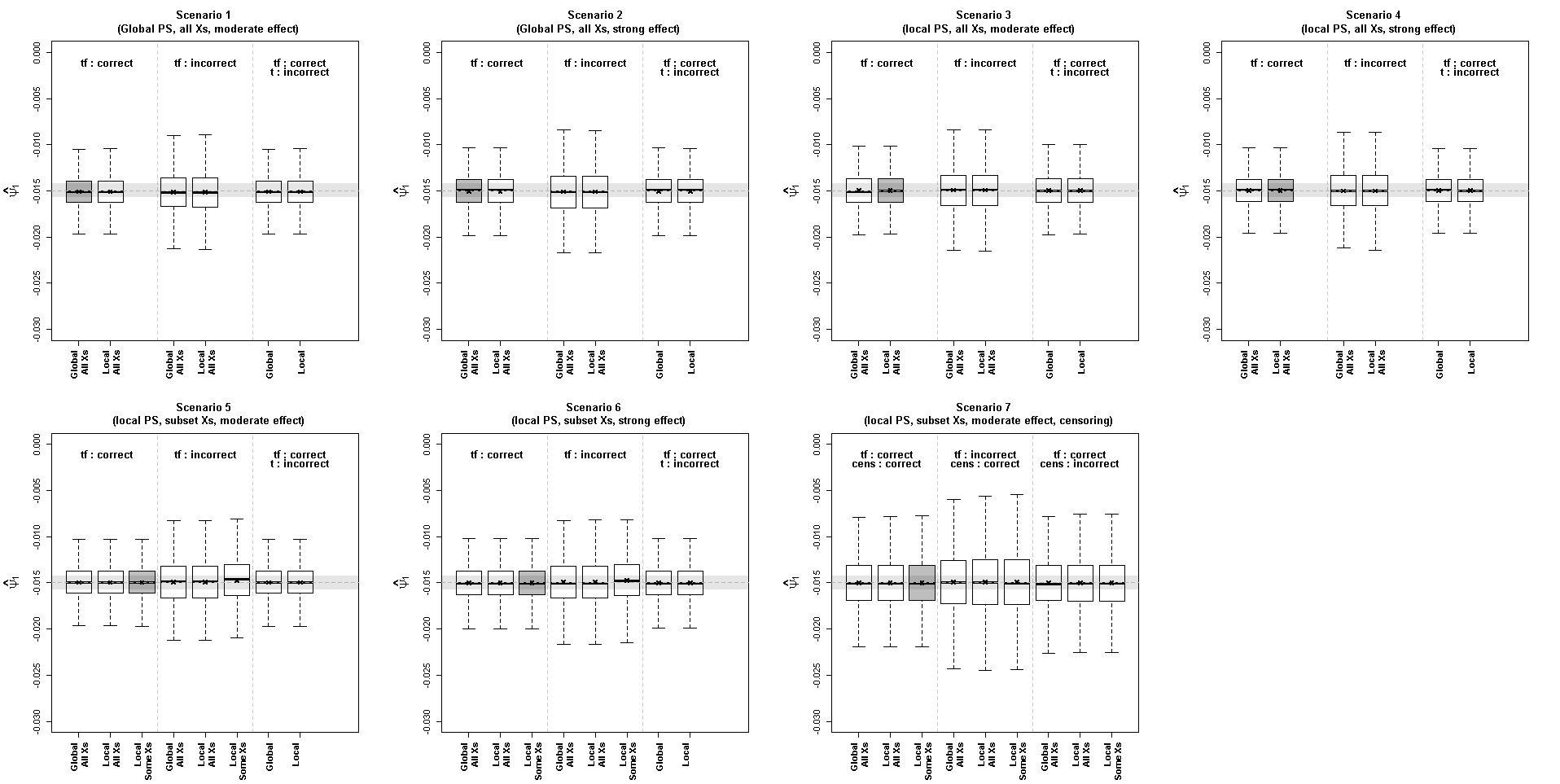}
        \caption{Simulation results: large sample size and small treatment effect settings - Performance of the methods over 1000 iterations in the estimation of $\psi_{1}$. For each scenario, the grey filling indicates the configuration under which the data have been simulated and the light-grey band represents the interval in which the absolute relative bias is less than or equal to 5\%. `PS' denotes propensity score, `tf' denotes treatment-free model, and 'cens' denotes censoring model.}
    \end{center}
\end{figure}
\end{landscape}

\begin{table}\label{tab:SmallNLowTmt}
{
\caption{Simulation results: Simulations with a large sample size and large treatment effect: Performance of the methods over 1000 iterations: mean estimate (95\%CI), relative bias (\%, denoted RB) and root mean squared error (RMSE) of $\hat\psi_0$, and difference in value function (dVF) between true and estimated ITR with its standard error (SE). Sc.~denotes scenario. }
\begin{tabular}{lllllll}
\hline
{Method} & Sc. & {Mean [CI]} & {RB}  & {RMSE} &  & dVF (SE) \\
\hline
&  & \multicolumn{5}{c}{\scriptsize{\textbf{Treatment-free model correctly specified, propensity score}}}\\
&  & \multicolumn{5}{c}{\scriptsize{\textbf{relies on all or a subset of all possible confounders}}} \\
\hline
Global All $X$s & 1 & 4.000 [3.966;4.034] & -0.010 & $2.90\times10^{-4}$ & & $-3.00\times10^{-6}(4.20\times10^{-6})$\rule{0pt}{8pt}\\
Local All $X$s & 1 & 4.000 [3.951;4.048] & -0.009 & $2.90\times10^{-4}$ & & $-3.00\times10^{-6}(4.20\times10^{-6})$\rule{0pt}{8pt}\\

\hline

Global All $X$s & 2 & 4.000 [3.966;4.034] & $<0.001$ & $2.90\times10^{-4}$ & & $-3.20\times10^{-6}(4.40\times10^{-6})$\rule{0pt}{8pt}\\
Local All $X$s & 2 & 4.000 [3.952;4.048] & $<0.001$ & $2.90\times10^{-4}$ & & $-3.20\times10^{-6}(4.40\times10^{-6})$\rule{0pt}{8pt}\\

\hline

Global All $X$s & 3 & 4.000 [3.966;4.034] & -0.007 & $2.80\times10^{-4}$ & & $-2.80\times10^{-6}(4.30\times10^{-6})$\rule{0pt}{8pt}\\
Local All $X$s & 3 & 4.000 [3.951;4.048] & -0.007 & $2.80\times10^{-4}$ & & $-2.80\times10^{-6}(4.30\times10^{-6})$\rule{0pt}{8pt}\\

\hline

Global All $X$s & 4 & 4.000 [3.966;4.034] & -0.003 & $2.50\times10^{-4}$ & & $-2.70\times10^{-6}(3.80\times10^{-6})$\rule{0pt}{8pt}\\
Local All $X$s & 4 & 4.000 [3.951;4.048] & -0.003 & $2.50\times10^{-4}$ & & $-2.70\times10^{-6}(3.80\times10^{-6})$\rule{0pt}{8pt}\\

\hline

Global All $X$s & 5 &  3.999 [3.965;4.033] &-0.016  & $3.00\times10^{-4}$ & & $-3.00\times10^{-6}(4.20\times10^{-6})$\rule{0pt}{8pt}\\
Local All $X$s & 5 & 3.999 [3.951;4.048] & -0.016 &$3.00\times10^{-4}$  & & $-3.00\times10^{-6}(4.20\times10^{-6})$\rule{0pt}{8pt}\\
Local Some $X$s & 5 &3.999 [3.951;4.048]  & -0.016 & $3.00\times10^{-4}$ & & $-3.00\times10^{-6}(4.20\times10^{-6})$\rule{0pt}{8pt}\\

\hline

Global All $X$s & 6 & 4.000 [3.966;4.034] &-0.009  & $2.80\times10^{-4}$ & & $-3.10\times10^{-6}(4.40\times10^{-6})$\rule{0pt}{8pt}\\
Local All $X$s & 6 & 4.000 [3.951;4.048] & -0.009 & $2.80\times10^{-4}$ & & $-3.10\times10^{-6}(4.40\times10^{-6})$\rule{0pt}{8pt}\\
Local Some $X$s & 6 & 4.000 [3.951;4.048] & -0.009 & $2.80\times10^{-4}$ & & $-3.10\times10^{-6}(4.40\times10^{-6})$\rule{0pt}{8pt}\\

\hline

Global& 7 & 4.006 [3.502;4.510] & 0.154 & $6.30\times10^{-4}$   & & $-6.80\times10^{-6}(1.05\times10^{-5})$\rule{0pt}{8pt}\\
Local All $X$s  & 7 & 4.006 [3.515;4.497] & 0.151 & $6.80\times10^{-4}$ & & $-7.40\times10^{-6}(1.15\times10^{-5})$\rule{0pt}{8pt}\\
Local Some $X$s  & 7 & 4.007 [3.519;4.495]  & 0.176 & $6.80\times10^{-4}$ & & $-7.40\times10^{-6}(1.15\times10^{-5})$\rule{0pt}{8pt}\\

\hline
&  & \multicolumn{5}{c}{\scriptsize{\textbf{Treatment-free model misspecified, propensity score relies on  }}} \\
&  & \multicolumn{5}{c}{\scriptsize{\textbf{all or a subset of all possible confounders}}} \\
\hline
Global All $X$s & 1 & 3.999 [3.955;4.043] & -0.015 & $5.30\times10^{-4}$ & & $-4.10\times10^{-6}(6.00\times10^{-6})$\rule{0pt}{8pt}\\
Local All $X$s & 1 & 3.999 [3.937;4.062] & -0.015 & $5.30\times10^{-4}$ & & $-4.10\times10^{-6}(6.00\times10^{-6})$\rule{0pt}{8pt}\\

\hline

Global All $X$s & 2 & 4.000 [3.956;4.044] & -0.007 & $5.20\times10^{-4}$ & & $-4.10\times10^{-6}(5.70\times10^{-6})$\rule{0pt}{8pt}\\
Local All $X$s & 2 & 4.000 [3.937;4.062] & -0.007 & $5.20\times10^{-4}$ & & $-4.10\times10^{-6}(5.70\times10^{-6})$\rule{0pt}{8pt}\\

\hline

Global All $X$s & 3 & 4.000 [3.956;4.044] & -0.007 & $5.20\times10^{-4}$ & & $-3.90\times10^{-6}(5.80\times10^{-6})$\rule{0pt}{8pt}\\
Local All $X$s & 3 & 4.000 [3.937;4.062] & -0.008 & $5.20\times10^{-4}$ & & $-3.80\times10^{-6}(5.80\times10^{-6})$\rule{0pt}{8pt}\\

\hline

Global All $X$s & 4 & 4.000 [3.956;4.044] & -0.001 & $4.50\times10^{-4}$ & & $-3.60\times10^{-6}(5.10\times10^{-6})$\rule{0pt}{8pt}\\
Local All $X$s & 4 & 4.000 [3.938;4.062] & -0.001 & $4.50\times10^{-4}$ & & $-3.60\times10^{-6}(5.10\times10^{-6})$\rule{0pt}{8pt}\\

\hline

Global All $X$s & 5 & 4.000 [3.956;4.043] & -0.011 & $5.30\times10^{-4}$ & & $-3.90\times10^{-6}(5.40\times10^{-6})$\rule{0pt}{8pt}\\
Local All $X$s & 5 & 4.000 [3.937;4.062] & -0.011 & $5.40\times10^{-4}$ &  & $-3.90\times10^{-6}(5.40\times10^{-6})$\rule{0pt}{8pt}\\
Local Some $X$s & 5 &3.997 [3.935;4.060]  & -0.063 & $5.40\times10^{-4}$ & & $-3.90\times10^{-6}(5.50\times10^{-6})$\rule{0pt}{8pt}\\

\hline

Global All $X$s & 6 & 3.999 [3.955;4.043] & -0.021 & $5.10\times10^{-4}$ & & $-4.20\times10^{-6}(5.80\times10^{-6})$\rule{0pt}{8pt}\\
Local All $X$s & 6 & 3.999 [3.937;4.061] & -0.021 & $5.10\times10^{-4}$ & & $-4.20\times10^{-6}(5.70\times10^{-6})$\rule{0pt}{8pt}\\
Local Some $X$s & 6 & 3.997 [3.935;4.059] & -0.073 & $5.20\times10^{-4}$ & & $-4.20\times10^{-6}(5.90\times10^{-6})$\rule{0pt}{8pt}\\

\hline

Global& 7 & 3.999 [3.364;4.634] & -0.016 & $1.04\times10^{-3}$   & & $-9.00\times10^{-6}(1.38\times10^{-5})$\rule{0pt}{8pt}\\
Local All $X$s  & 7 & 4.002 [3.371;4.632] & 0.040 & $1.14\times10^{-3}$  & & $-9.80\times10^{-6}(1.49\times10^{-5})$\rule{0pt}{8pt}\\
Local Some $X$s  & 7 & 4.002 [3.375;4.629] &  0.046 & $1.14\times10^{-3}$   & & $-9.80\times10^{-6}(1.49\times10^{-5})$\rule{0pt}{8pt}\\

\hline
&  & \multicolumn{5}{c}{\scriptsize{\textbf{Treatment-free model correctly specified, propensity score or }}} \\
&  & \multicolumn{5}{c}{\scriptsize{\textbf{censoring models misspecified (assumed independent of $X$s)}}} \\
\hline
Global  & 1 & 4.000 [3.965;4.034] &-0.010  & $2.90\times10^{-4}$  & & $-3.00\times10^{-6}(4.20\times10^{-6})$\rule{0pt}{8pt}\\
Local  & 1 & 4.000 [3.951;4.048] & -0.010 & $2.90\times10^{-4}$ & & $-3.00\times10^{-6}(4.20\times10^{-6})$\rule{0pt}{8pt}\\

\hline

Global  & 2 & 4.000 [3.966;4.034] & $<0.001$ &$2.90\times10^{-4}$  & & $-3.20\times10^{-6}(4.40\times10^{-6})$\rule{0pt}{8pt}\\
Local  & 2 & 4.000 [3.951;4.049] &  $<0.001$& $2.90\times10^{-4}$ & & $-3.20\times10^{-6}(4.40\times10^{-6})$\rule{0pt}{8pt}\\

\hline

Global  & 3 & 4.000 [3.965;4.034] &-0.007  & $2.80\times10^{-4}$ & & $-2.90\times10^{-6}(4.30\times10^{-6})$\rule{0pt}{8pt}\\
Local  & 3 & 4.000 [3.951;4.048] & -0.007 & $2.80\times10^{-4}$ & & $-2.80\times10^{-6}(4.30\times10^{-6})$\rule{0pt}{8pt}\\

\hline

Global & 4 & 4.000 [3.966;4.034] & -0.003 & $2.50\times10^{-4}$ & & $-2.70\times10^{-6}(3.80\times10^{-6})$\rule{0pt}{8pt}\\
Local  & 4 & 4.000 [3.951;4.048] & -0.003 & $2.50\times10^{-4}$ & & $-2.70\times10^{-6}(3.80\times10^{-6})$\rule{0pt}{8pt}\\

\hline

Global  & 5 & 3.999 [3.965;4.034] & -0.016 &  $3.00\times10^{-4}$ & & $-3.00\times10^{-6}(4.20\times10^{-6})$\rule{0pt}{8pt}\\
Local  & 5 & 3.999 [3.951;4.048] & -0.016 &  $3.00\times10^{-4}$ & & $-3.00\times10^{-6}(4.20\times10^{-6})$\rule{0pt}{8pt}\\

\hline

Global  & 6 & 4.000 [3.965;4.034] & -0.009 & $2.80\times10^{-4}$ & & $-3.10\times10^{-6}(4.40\times10^{-6})$\rule{0pt}{8pt}\\
Local  & 6 & 4.000 [3.951;4.048] & -0.009 &  $2.80\times10^{-4}$ & & $-3.10\times10^{-6}(4.40\times10^{-6})$\rule{0pt}{8pt}\\

\hline

Global& 7 & 4.006 [3.503;4.509] & 0.150 & $6.30\times10^{-4}$   & & $-6.80\times10^{-6}(1.04\times10^{-5})$\rule{0pt}{8pt}\\
Local All $X$s  & 7 & 4.005 [3.514;4.496] & 0.127 & $6.60\times10^{-4}$ & &$-7.20\times10^{-6}(1.10\times10^{-5})$\rule{0pt}{8pt}\\
Local Some $X$s  & 7 & 4.006 [3.517;4.495] & 0.155 & $6.60\times10^{-4}$ & & $-7.20\times10^{-6}(1.10\times10^{-5})$\rule{0pt}{8pt}\\
\hline
\end{tabular}
}

\parbox{18.5cm}{ ``Global All $X$s'' = Global propensity score with all possible confounders}
\parbox{18.5cm}{ ``Local All $X$s'' = Local propensity score with all possible confounders}
\parbox{18.5cm}{ ``Local Some $X$s'' = Local propensity score with known local confounders}
\parbox{18.5cm}{ ``Global'' = Global propensity score with treatment assumed to be randomly allocated}
\parbox{18.5cm}{ ``Local'' = Local propensity score with treatment assumed to be randomly allocated}
\end{table}

\begin{landscape}
\begin{figure}
   \begin{center}
        \includegraphics[scale=0.28]{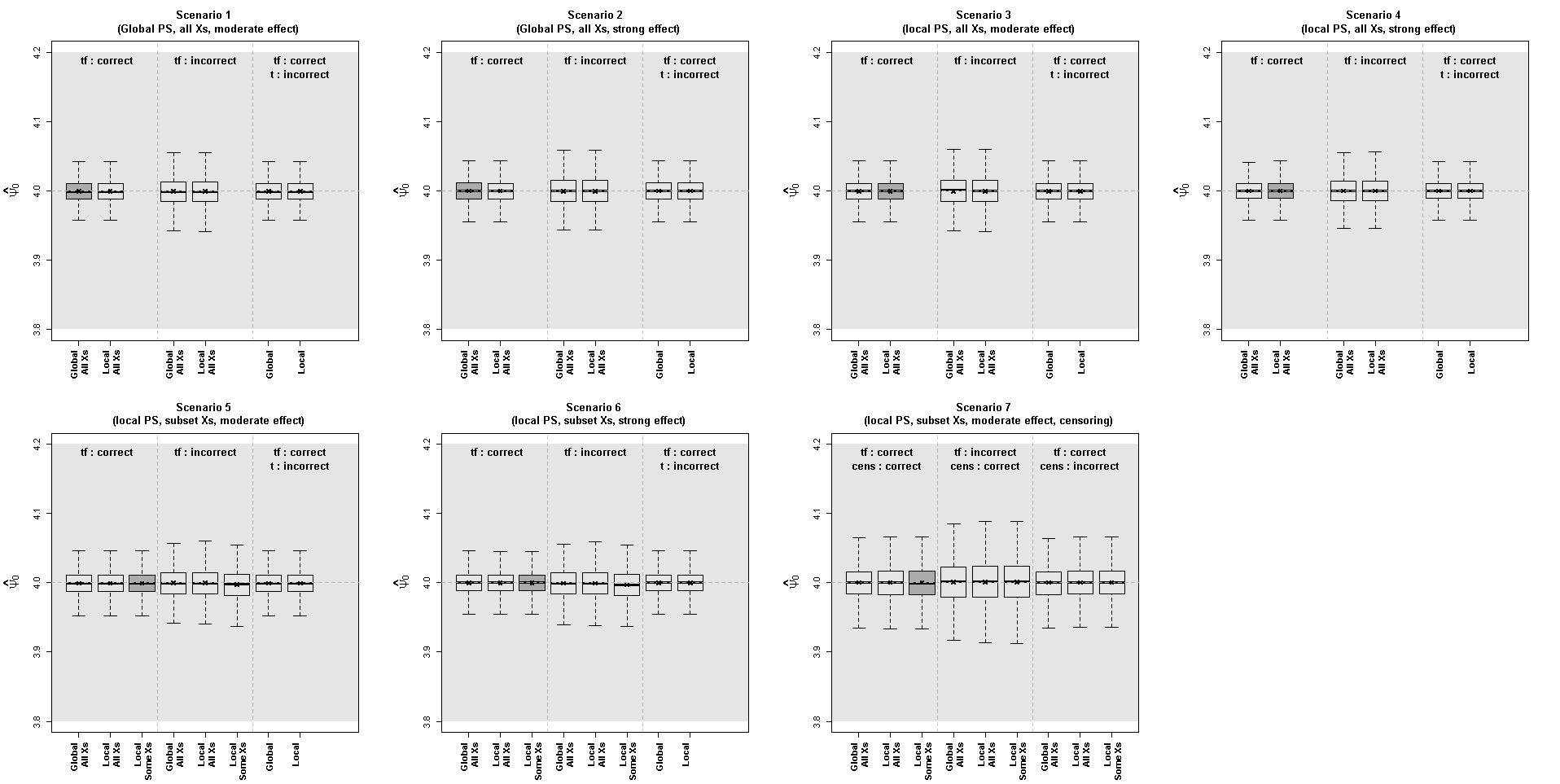}
        \caption{Simulation results: large sample size and large treatment effect settings - Performance of the methods over 1000 iterations in the estimation of $\psi_{0}$. For each scenario, the grey filling indicates the configuration under which the data have been simulated and the light-grey band represents the interval in which the absolute relative bias is less than or equal to 5\%. `PS' denotes propensity score, `tf' denotes treatment-free model, and 'cens' denotes censoring model.}
    \end{center}
\end{figure}
\end{landscape}

\begin{table}\label{tab:SmallNLowTmt}
{
\caption{Simulation results: Simulations with a large sample size and large treatment effect: Performance of the methods over 1000 iterations: mean estimate (95\%CI), relative bias (\%, denoted RB) and root mean squared error (RMSE) of $\hat\psi_1$, and difference in value function (dVF) between true and estimated ITR with its standard error (SE). Sc.~denotes scenario. }
\begin{tabular}{llllllll}
\hline
{Method} & Sc. &&&& {Mean [CI]} & {RB}  & {RMSE} \\
\hline
&  & \multicolumn{6}{c}{\scriptsize{\textbf{Treatment-free model correctly specified, propensity score}}}\\
&  & \multicolumn{6}{c}{\scriptsize{\textbf{relies on all or a subset of all possible confounders}}} \\
\hline
Global All $X$s & 1 &&&& -0.550 [-0.554;-0.546] &  -0.005 & $3.40\times10^{-6}$ \rule{0pt}{8pt}\\
Local All $X$s & 1 &&&& -0.550 [-0.555;-0.545] &  -0.004 & $3.40\times10^{-6}$ \rule{0pt}{8pt}\\

\hline

Global All $X$s & 2 &&&& -0.550 [-0.554;-0.546] & -0.004 & $3.40\times10^{-6}$ \rule{0pt}{8pt}\\
Local All $X$s & 2 &&&& -0.550 [-0.555;-0.545] & -0.004 & $3.40\times10^{-6}$ \rule{0pt}{8pt}\\

\hline

Global All $X$s & 3 &&&& -0.550 [-0.554;-0.546] & -0.010 & $3.40\times10^{-6}$ \rule{0pt}{8pt}\\
Local All $X$s & 3 &&&& -0.550 [-0.555;-0.545] & -0.010 & $3.40\times10^{-6}$\rule{0pt}{8pt}\\

\hline

Global All $X$s & 4 &&&& -0.550 [-0.554;-0.546] & 0.002 & $2.90\times10^{-6}$ \rule{0pt}{8pt}\\
Local All $X$s & 4 &&&& -0.550 [-0.555;-0.545] & 0.002 & $2.90\times10^{-6}$ \rule{0pt}{8pt}\\

\hline

Global All $X$s & 5 &&&&  -0.550 [-0.554;-0.546] &-0.013  & $3.50\times10^{-6}$ \rule{0pt}{8pt}\\
Local All $X$s & 5 &&&& -0.550 [-0.555;-0.545] & -0.013 &$3.50\times10^{-6}$ \rule{0pt}{8pt}\\
Local Some $X$s & 5 &&&&-0.550 [-0.555;-0.545]  & -0.013 & $3.50\times10^{-6}$ \rule{0pt}{8pt}\\

\hline

Global All $X$s & 6 &&&& -0.550 [-0.554;-0.546] & -0.005 & $3.30\times10^{-6}$ \rule{0pt}{8pt}\\
Local All $X$s & 6 &&&& -0.550 [-0.555;-0.545] & -0.005 & $3.30\times10^{-6}$ \rule{0pt}{8pt}\\
Local Some $X$s & 6 &&&& -0.550 [-0.555;-0.545] & -0.005 & $3.30\times10^{-6}$ \rule{0pt}{8pt}\\

\hline

Global& 7 &&&&  -0.551 [-0.604;-0.497] & 0.103 & $7.00\times10^{-6}$ \rule{0pt}{8pt}\\
Local All $X$s  & 7 &&&&  -0.551 [-0.604;-0.497] & 0.109 & $7.50\times10^{-6}$ \rule{0pt}{8pt}\\
Local Some $X$s  & 7 &&&&  -0.551 [-0.604;-0.498] & 0.122 & $7.50\times10^{-6}$ \rule{0pt}{8pt}\\

\hline
&  & \multicolumn{6}{c}{\scriptsize{\textbf{Treatment-free model misspecified, propensity score relies on  }}} \\
&  & \multicolumn{6}{c}{\scriptsize{\textbf{all or a subset of all possible confounders}}} \\
\hline
Global All $X$s & 1 &&&& -0.550 [-0.555;-0.545] &  -0.009 & $6.40\times10^{-6}$ \rule{0pt}{8pt}\\
Local All $X$s & 1 &&&& -0.550 [-0.557;-0.543] &  -0.009 & $6.40\times10^{-6}$ \rule{0pt}{8pt}\\

\hline

Global All $X$s & 2 &&&& -0.550 [-0.555;-0.545] & -0.010 & $6.20\times10^{-6}$ \rule{0pt}{8pt}\\
Local All $X$s & 2 &&&& -0.550 [-0.557;-0.543] & -0.010 & $6.30\times10^{-6}$ \rule{0pt}{8pt}\\

\hline

Global All $X$s & 3 &&&& -0.550 [-0.555;-0.545] & -0.010 & $6.30\times10^{-6}$ \rule{0pt}{8pt}\\
Local All $X$s & 3 &&&&-0.550 [-0.557;-0.543] & -0.010 & $6.30\times10^{-6}$ \rule{0pt}{8pt}\\

\hline

Global All $X$s & 4 &&&& -0.550 [-0.555;-0.545]  & 0.004 & $5.40\times10^{-6}$ \rule{0pt}{8pt}\\
Local All $X$s & 4 &&&& -0.550 [-0.557;-0.543] & 0.004 & $5.40\times10^{-6}$ \rule{0pt}{8pt}\\

\hline

Global All $X$s & 5 &&&& -0.550 [-0.555;-0.545] & -0.009 &  $6.40\times10^{-6}$\rule{0pt}{8pt}\\
Local All $X$s & 5 &&&& -0.550 [-0.557;-0.543] & -0.009 &  $6.40\times10^{-6}$ \rule{0pt}{8pt}\\
Local Some $X$s & 5 &&&&-0.550 [-0.557;-0.543]  & -0.050 & $6.50\times10^{-6}$ \rule{0pt}{8pt}\\

\hline

Global All $X$s & 6 &&&& -0.550 [-0.555;-0.545] & -0.014 & $6.10\times10^{-6}$ \rule{0pt}{8pt}\\
Local All $X$s & 6 &&&& -0.550 [-0.557;-0.543] & -0.014 & $6.10\times10^{-6}$ \rule{0pt}{8pt}\\
Local Some $X$s & 6 &&&& -0.550 [-0.557;-0.543] & -0.055 & $6.20\times10^{-6}$ \rule{0pt}{8pt}\\

\hline

Global& 7 &&&&  -0.550 [-0.618;-0.482] &  -0.030 & $1.18\times10^{-5}$ \rule{0pt}{8pt}\\
Local All $X$s  & 7 &&&&  -0.550 [-0.619;-0.481] & 0.018 & $1.30\times10^{-5}$ \rule{0pt}{8pt}\\
Local Some $X$s  & 7 &&&&  -0.550 [-0.619;-0.481] & 0.017 & $1.30\times10^{-5}$ \rule{0pt}{8pt}\\

\hline
&  & \multicolumn{6}{c}{\scriptsize{\textbf{Treatment-free model correctly specified, propensity score or }}} \\
&  & \multicolumn{6}{c}{\scriptsize{\textbf{censoring models misspecified (assumed independent of $X$s)}}} \\
\hline
Global  & 1 &&&& -0.550 [-0.554;-0.546] & -0.005 & $3.40\times10^{-6}$ \rule{0pt}{8pt}\\
Local  & 1 &&&& -0.550 [-0.555;-0.545] & -0.005 & $3.40\times10^{-6}$ \rule{0pt}{8pt}\\

\hline

Global  & 2 &&&& -0.550 [-0.554;-0.546] & -0.004 & $3.40\times10^{-6}$ \rule{0pt}{8pt}\\
Local  & 2 &&&& -0.550 [-0.555;-0.545] & -0.004 & $3.40\times10^{-6}$ \rule{0pt}{8pt}\\

\hline

Global  & 3 &&&& -0.550 [-0.554;-0.546] & -0.010 & $3.40\times10^{-6}$ \rule{0pt}{8pt}\\
Local  & 3 &&&& -0.550 [-0.555;-0.545] & -0.010 & $3.40\times10^{-6}$ \rule{0pt}{8pt}\\

\hline

Global & 4 &&&& -0.550 [-0.554;-0.546] & 0.002 & $2.90\times10^{-6}$ \rule{0pt}{8pt}\\
Local  & 4 &&&& -0.550 [-0.555;-0.545] & 0.002 & $2.90\times10^{-6}$ \rule{0pt}{8pt}\\

\hline

Global  & 5 &&&& -0.550 [-0.554;-0.546] & -0.013 & $3.50\times10^{-6}$ \rule{0pt}{8pt}\\
Local  & 5 &&&& -0.550 [-0.555;-0.545] & -0.013 & $3.50\times10^{-6}$ \rule{0pt}{8pt}\\

\hline

Global  & 6 &&&& -0.550 [-0.554;-0.546] & -0.005 & $3.30\times10^{-6}$ \rule{0pt}{8pt}\\
Local  & 6 &&&& -0.550 [-0.555;-0.545] & -0.005 & $3.30\times10^{-6}$ \rule{0pt}{8pt}\\

\hline

Global& 7 &&&& -0.551 [-0.604;-0.497] &  0.101 & $7.00\times10^{-6}$ \rule{0pt}{8pt}\\
Local All $X$s  & 7 &&&& -0.551 [-0.604;-0.497] & 0.097 & $7.40\times10^{-6}$ \rule{0pt}{8pt}\\
Local Some $X$s  & 7 &&&& -0.551 [-0.604;-0.498] & 0.113 & $7.40\times10^{-6}$ \rule{0pt}{8pt}\\
\hline 
\end{tabular}
}

\parbox{18.5cm}{ ``Global All $X$s'' = Global propensity score with all possible confounders}
\parbox{18.5cm}{ ``Local All $X$s'' = Local propensity score with all possible confounders}
\parbox{18.5cm}{ ``Local Some $X$s'' = Local propensity score with known local confounders}
\parbox{18.5cm}{ ``Global'' = Global propensity score with treatment assumed to be randomly allocated}
\parbox{18.5cm}{ ``Local'' = Local propensity score with treatment assumed to be randomly allocated}
\end{table}

\begin{landscape}
\begin{figure}
   \begin{center}
        \includegraphics[scale=0.28]{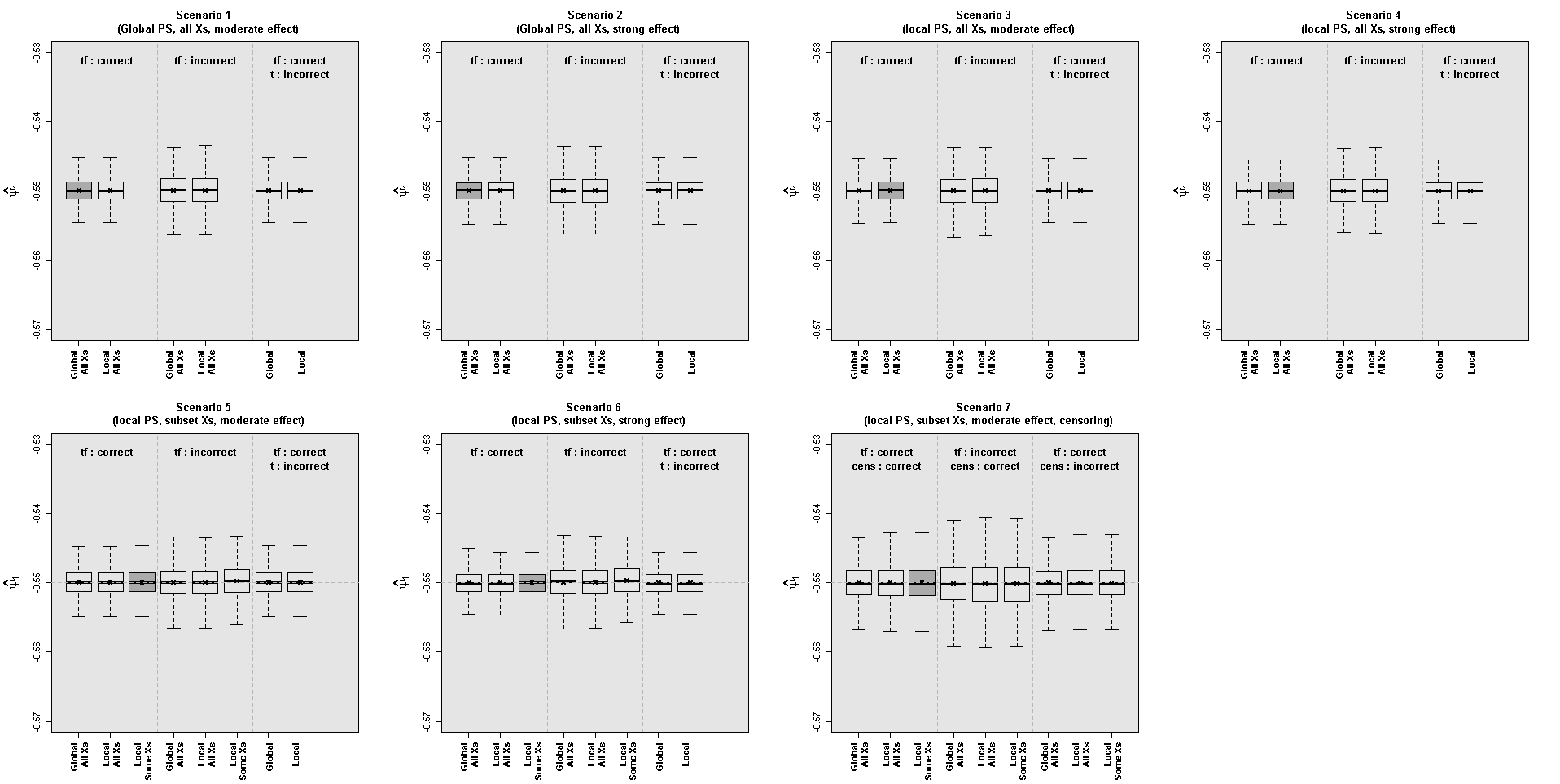}
        \caption{Simulation results: large sample size and large treatment effect settings - Performance of the methods over 1000 iterations in the estimation of $\psi_{1}$. For each scenario, the grey filling indicates the configuration under which the data have been simulated and the light-grey band represents the interval in which the absolute relative bias is less than or equal to 5\%. `PS' denotes propensity score, `tf' denotes treatment-free model, and 'cens' denotes censoring model.}
    \end{center}
\end{figure}
\end{landscape}
\newpage

\FloatBarrier
\small

\section*{C. Conservative variance formula}

Recall from Section 2.3 of the main paper that the doubly robust distributed regression estimator from $J$ centres can be expressed as
\[
\hat{\boldsymbol{\theta}} = \left(\sum_{j=1}^J\boldsymbol{X}_j{^T}W_j\boldsymbol{X}_j\right)^{-1}\left(\sum_{j=1}^J\boldsymbol{X}_j{^T}W_j\tilde{Y}_j,
\right) \]
where $\boldsymbol{X}_j$, $W_j$, and $\tilde{Y}_j$ are, respectively, the centre-specific design matrix, diagonal weight matrix, and vector of log-transformed outcomes for each of the $j=1,...,J$ centres. Let $\hat\sigma_j^2 = (n_j - p)^{-1}R_j{^T}W_jR_j$ for $n_j$ the sample size at centre $j$, $p$ the number of columns in $\boldsymbol{X}_j$, and $R_j$ the vector of residuals from centre $j$. Then the variance of $\hat{\boldsymbol{\theta}}$ is given by\footnote{Hedges LV, Tipton E, Johnson MC (2010) Robust variance estimation in meta-regression with dependent effect size estimates. \textit{Research Synthesis Methods} \textbf{1}:39–65.}
\[
\left(\sum_{j=1}^J\boldsymbol{X}_j{^T}W_j\boldsymbol{X}_j\right)^{-1}
\left(\sum_{j=1}^J\boldsymbol{X}_j{^T}W_j\Sigma_jW_j\boldsymbol{X}_j\right)
\left(\sum_{j=1}^J\boldsymbol{X}_j{^T}W_j\boldsymbol{X}_j\right)^{-1},
\]
where $\Sigma_j$ is a diagonal matrix whose entries equal $\hat\sigma_jw_{ji}^{-1/2}$ for $w_{ji}$ the $i$th element of the diagonal matrix $W_j$.

This approximation performs well in practice, underestimating the standard error by as little as 2.8\% in Scenario 5 when the treatment-free model is correctly specified, and up to 5.2\% in Scenario 7 when the propensity score is misspecified.
\newpage
\FloatBarrier

\section*{D. Additional CPRD results}

\FloatBarrier


\begin{table}
\caption{Proportion of patients in the study cohort in each region (presented in an ascending order), Clinical Practice Research Datalink, United Kingdom, 1998-2017}
\label{tabregion}
\begin{tabular}{lc}
\hline\noalign{\smallskip}
Region  &  Number of patients (\%) \\ \hline
North East& 4841 (2.0)\\
Yorkshire and the Humber& 9595 (3.9)\\
East Midlands& 10,756 (4.4)\\
London & 25,952 (10.5) \\
East of England & 27,430 (11.1)\\
South East Coast & 31,612 (12.8) \\
West Midlands & 31,711 (12.9) \\
South Central & 32,057 (13.0)\\
South West & 32,454 (13.2) \\
North West& 40,095 (16.3)\\
\noalign{\smallskip}\hline
\end{tabular}
\end{table}

\begin{figure*}[htp]
   \includegraphics[width=12cm]{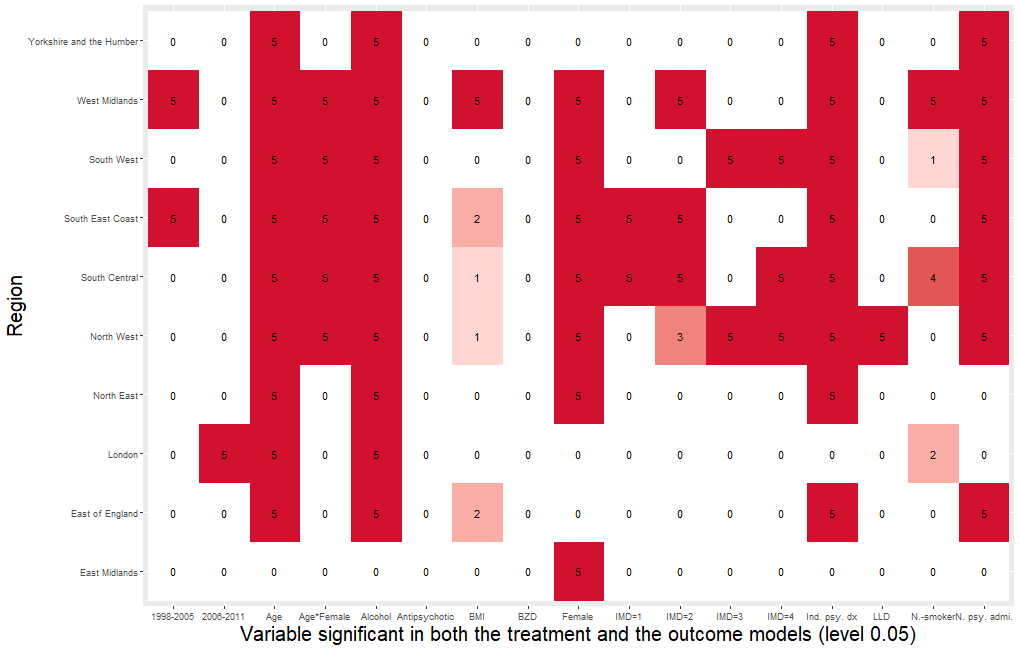}
\caption{Frequency of selection of each variable (level of significance of 0.05) over a total of 5 imputed datasets, for each region separately. Variables had to be statistically significant (p-value$<$0.05) in both respective logistic and Cox univariate models predicting the treatment and the outcome to be counted in the heat map. Each box contains the number (0 to 5) of datasets in which the variable's coefficient was found to be statistically significant, and the map is colored in accordance with that number, with white corresponding to 0 such dataset, and bright red corresponding to 5 datasets, Clinical Practice Research Datalink, United Kingdom, 1998-2017}
\label{heatmap:trmt}
\end{figure*}

\begin{figure*}[htp]
   \includegraphics[width=12cm]{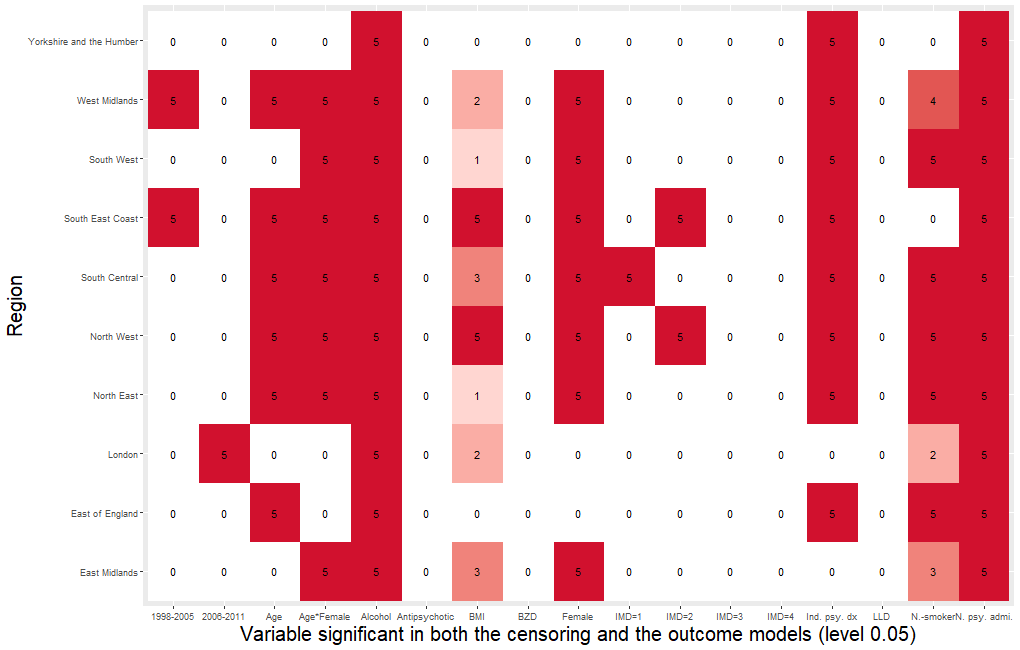}
\caption{Frequency of selection of each variable (level of significance of 0.05) over a total of 5 imputed datasets, for each region separately.  Variables had to be statistically significant (p-value$<$0.05) in both respective logistic and Cox univariate models predicting censoring and the outcome to be counted in the heat map. Each box contains the number (0 to 5) of datasets in which the variable's coefficient was found to be statistically significant, and the map is colored in accordance with that number, with white corresponding to 0 such dataset, and bright red corresponding to 5 datasets, Clinical Practice Research Datalink, United Kingdom, 1998-2017}
\label{heatmap:censor}
\end{figure*}

\bibliographystyle{chicago}       
\bibliography{biblioPrivacyDWSurv}   

\end{document}